\documentclass[aps,twocolumn,pra,10pt,superscriptaddress,nofootinbib]{revtex4-2}

\usepackage{comment}
\usepackage[T1]{fontenc}
\usepackage{dsfont}
\usepackage{amsfonts}
\usepackage{amsmath}
\usepackage{bm}
\usepackage{amssymb}
\usepackage{graphicx}
\usepackage{braket}
\usepackage[table]{xcolor}
\usepackage{colortbl}
\usepackage{multirow} 
\usepackage{booktabs}
\usepackage[normalem]{ulem} 

\renewcommand{\leq}{\leqslant}
\renewcommand{\geq}{\geqslant}

\usepackage[colorlinks=true,urlcolor=blue,citecolor=blue,linkcolor=blue]{hyperref}
\DeclareGraphicsExtensions{.eps,.ps,.pdf}
\usepackage{tikz}
\usepackage{array}

\definecolor{myblue}{RGB}{69, 104, 134}
\newcommand{\bcolor}{\cellcolor{myblue!20}}

\definecolor{rowgray}{gray}{0.9}
\definecolor{rowblue}{RGB}{230, 240, 255}
\definecolor{rowgreen}{RGB}{230, 255, 230}

\newcommand{\la}{\lambda}
\newcommand{\si}{\sigma}

\newcommand{\bmu}{\boldsymbol{\mu}}
\newcommand{\bfo}{\mbox{\boldmath $0$}}

\newcommand{\PP}{\hat{\mathcal{P}}}
\newcommand{\QQ}{\hat{\mathcal{Q}}}

\newcommand{\RR}{\mathcal{R}}

\newcommand{\Scal}{\mathcal{S}}

\def \Tr {\text{Tr}}

\newtheorem{theorem}{Theorem}

\newtheorem{prop}{Proposition}
\newtheorem{lemma}{Lemma}
\newtheorem{conjecture}{Conjecture}

\newcommand{\Hs}{\mathcal{H}}

\newcommand{\Sh}{C}
\newcommand{\Shpsi}{S^{(e)}}
\newcommand{\wt}{\mathrm{wt}}

\newcommand*{\mbinom}[2]{\scalebox{1.3}{\ensuremath{\binom{#1}{#2}}}}

\begin{document} 
\title{Multiqubit monogamy relations beyond shadow inequalities} 
\newcommand{\majulab}{MajuLab, CNRS–UCA-SU-NUS-NTU International Joint Research Laboratory}
\newcommand{\cqt}{Centre for Quantum Technologies, National University of Singapore, 17543 Singapore, Singapore}
\newcommand{\LPTMS}{Université Paris-Saclay, CNRS, LPTMS, 91405, Orsay, France}

\author{Eduardo Serrano-Ens\'astiga}
\email{ed.ensastiga@uliege.be}
\affiliation{Institut de Physique Nucléaire, Atomique et de Spectroscopie, CESAM, University of Liège
\\
B-4000 Liège, Belgium}
\author{Olivier Giraud}
\email{olivier.giraud@universite-paris-saclay.fr}
\affiliation{\LPTMS}
\affiliation{\majulab}
\affiliation{\cqt}
\author{John Martin}
\email{jmartin@uliege.be}
\affiliation{Institut de Physique Nucléaire, Atomique et de Spectroscopie, CESAM, University of Liège
\\
B-4000 Liège, Belgium}

\date{January 7, 2026}
\begin{abstract}
Multipartite quantum systems are subject to monogamy relations that impose fundamental constraints on the distribution of quantum correlations between subsystems. These constraints can be studied quantitatively through sector lengths, defined as the average value of $m$-body correlations, which have applications in quantum information theory and coding theory. In this work, we derive a set of monogamy inequalities that complement the shadow inequalities, enabling a complete characterization of the numerical range of sector lengths for systems with $N\leq 5$ qubits in a pure state. This range forms a convex polytope, facilitating the efficient extremization of key physical quantities, such as the linear entropy of entanglement and the quantum shadow enumerators, by a simple evaluation at the polytope vertices. For larger systems ($N\geq 6$), we highlight a significant increase in complexity that neither our inequalities nor the shadow inequalities can fully capture.
\end{abstract}

\maketitle
\section{Introduction}
The constituents of a multipartite quantum system exhibit correlations, some of which arise solely from their quantum nature. These correlations include entanglement, which has no classical analog. However, quantum correlations are not arbitrary but are subject to fundamental physical constraints. A well-known example is the monogamy of entanglement in three-qubit systems~\cite{PhysRevA.60.4344,PhysRevA.61.052306}, which restricts how entanglement is shared among three qubits, thereby limiting the ability of each qubit to be entangled with more than one other. More generally, multipartite quantum systems obey similar monogamy relations that limit the distribution of quantum correlations between subsystems. These correlations are essential resources for applications in quantum metrology and quantum information processing. It is therefore very interesting to characterize and quantify them, assuming only that the global state is pure.

For multiqubit systems, a general state $\rho$ is fully specified by the expectation values of Pauli strings,
\begin{equation}
\label{Eq.Pauli.strings}
\si_{\bmu} = \si_{\mu_1 \mu_2 \dots \mu_N} = \sigma_{\mu_1} \otimes \dots \otimes \sigma_{\mu_N}\,,
\end{equation}
where $\sigma_0$ denotes the identity operator and $\sigma_a$ ($a = 1, 2, 3$) are the Pauli matrices, and each index $\mu_\alpha$ takes values in $\{0,1,2,3\}$. Constraints on quantum correlations are therefore encoded in the expectation values $\langle \si_{\bmu} \rangle$. Grouping these expectation values according to the number of non-identity Pauli operators forms the basis for defining the sector lengths $S_{m}$~\cite{Asc.Cal.Hei.Bri:04,PhysRevA.94.042302,wyderka2020characterizing}. These quantify the average correlations, both classical and quantum, between $m$ qubits in the $N$-qubit system. More precisely, they are defined as the average of the squared expectation values of the Pauli strings that act in a non-trivial way on the qubits. Although these quantities are far fewer in number than Pauli strings, they have many practical applications, including the detection of $m$-partite entanglement~\cite{PhysRevA.94.042302,HORODECKI1996377,PhysRevLett.108.240501,PhysRevA.98.052317,PhysRevA.98.052317,PhysRevLett.95.260502,PhysRevA.87.034301,eltschka2015monogamy,wyderka2020characterizing,PhysRevA.100.032307,9605348}, generalized uncertainty relations~\cite{10.1063/1.2943685,PhysRevA.107.062211,PhysRevLett.132.200202}, quantum secret sharing~\cite{PhysRevA.86.052335}, the quantum marginal problem~\cite{Huber_Severini_2018}, and noise robustness of entanglement~\cite{Miller_2023}, among others~\cite{CIESLINSKI20241}.

Sector lengths were originally introduced within the framework of quantum coding theory~\cite{PhysRevLett.78.1600}, where they are more commonly referred to as Shor-Laflamme enumerators~\cite{PhysRevA.69.052330}. In this context, $S_{m}$ quantifies the effect of errors acting simultaneously on $m$ qubits~\cite{PhysRevLett.77.2585,PhysRevA.56.1721,PhysRevLett.78.1600,681316,681315,Rains:796376,817508}. As such, the numerical values that $S_{m}$ can take determine the feasibility of constructing quantum codes, and, conversely, constraints on quantum codes impose bounds on the sector lengths. Several key bounds on quantum codes, and thus on sector lengths, include the quantum analogs of classical results such as the Hamming bound~\cite{PhysRevLett.77.2585}, the singleton bound~\cite{PhysRevA.56.1721}, and the MacWilliams identities~\cite{PhysRevLett.78.1600,Rains:796376}.  Additional constraints arise from quantum shadow enumerators~\cite{681315,Rains:796376}, whose physical interpretation and experimental measurement have been recently explored in~\cite{miller2024}. In particular, certain classes of highly entangled states, such as absolutely maximally entangled (AME) states~\cite{GISIN19981,HIGUCHI2000213,Brown_2005,Borras_2007,PhysRevLett.118.200502,Huber_2018,doi:10.1142/S0219749904000079,PhysRevA.86.052335} and $m$-uniform states~\cite{Borras_2007,PhysRevA.69.052330,PhysRevA.87.012319,PhysRevA.90.022316,PhysRevA.100.032112,Zhang.Pang.Shao.Zhu:24,zhang2024extremal,ning2025,Enrquez2016}, are associated with optimal quantum error-correcting codes~\cite{PhysRevA.55.900,PhysRevA.67.012313,PhysRevA.69.052330,doi:10.1142/S0219749904000079,Huber2020quantumcodesof}.

In this work, we derive multiqubit monogamy relations in the form of inequalities involving only sector lengths. By combining these inequalities with those already known, we obtain a set of constraints that, in certain cases, fully characterize the realizable values of sector lengths for pure states. For systems with up to five qubits, we show that the set of achievable sector lengths forms a polytope. Similar polytope structures have already been found for mixed states  of two and three qubits in Ref.~\cite{wyderka2020characterizing}. This complete characterization allows us to efficiently optimize various quantities related to entanglement and quantum correlations over all pure states with these numbers of qubits. Furthermore, we highlight several connections between sector lengths, linear entanglement entropy, and quantum shadow enumerators.

The paper is organized as follows. In Sec.~\ref{sec2:definitions}, we review basic monogamy relations involving sector lengths. In Sec.~\ref{Sec3.Monogamy}, we derive additional monogamy relations based on the time-reversed state. 
In Secs.~\ref{Sec4.Num.Range} and \ref{Sec5.infinite}, we examine the numerical values that sector lengths can take for systems with up to $N=8$ qubits and for arbitrary number of qubits in particular cases, respectively. We also explore their relevance to entanglement theory, quantum coding theory, and the verification of monogamy conjectures in Section~\ref{Sec6.App}. Finally, Section~\ref{Sec7.Conclusions} presents our conclusions and perspectives.
\section{Definitions and basic inequalities}
\label{sec2:definitions}
\subsection{Sector lengths and general inequalities}
\label{sec:additionalIneqs}
Consider an arbitrary $N$-qubit state $\rho$. We expand $\rho$ in the Pauli string basis~\eqref{Eq.Pauli.strings} as
\begin{equation}
\label{tensorrepmixed}
    \rho = \frac{1}{2^N} \sum_{\bmu} x_{\bmu} \sigma_{\bmu} \, ,
\end{equation}
where the coefficients $x_{\bmu} = \Tr \left( \rho\, \si_{\bmu}  \right)$ are real numbers. Throughout this work, we follow the convention that greek indices run from $0$ to $3$, while latin indices are restricted to the range $1$ to $3$. The \emph{weight} of a Pauli string, denoted by $\wt(\si_{\bm{\mu}})$, is defined as the number of nonzero indices in the multi-index $\bm{\mu}$~\cite{681316}. Accordingly, a correlation involving $m$ parties corresponds to a Pauli string with $\wt(\si_{\bm{\mu}})=m$.

Consider a subset $A\subset \{1 , \dots , N \}$ of $k$ qubits. We denote by $\rho_A$ the reduced density matrix obtained by tracing out the complementary set $\bar{A} = \{1, \dots, N\} \setminus A$, \emph{i.e.}, the qubits not in $A$. A useful property of the parametrization~\eqref{tensorrepmixed} is that the coefficients $x^A_{\mu_1  \ldots  \mu_k}$ of $\rho_A$ are directly inherited from those of $\rho$, with zeros inserted in place of the indices associated with $\bar{A}$. For example, when tracing out the last $N-k$ qubits, we have
$A= \{1 , \dots , k \}$ and 
\begin{equation}
\label{Eq.coord.reduced.mixed}
x^A_{\mu_1  \ldots  \mu_k} = x_{\mu_1  \ldots \mu_k 0 \ldots 0 } = x_{\mu_1  \ldots \mu_k \mathbf{0}_{N-k}}
\end{equation}
where $\mathbf{0}_{N-k}$ denotes a string of $N-k$ zeros~\cite{PhysRevLett.114.080401,PhysRevA.87.012319}.

The sector length $S_{m}$ is defined as the sum of the squared coefficients corresponding to $m$-body correlations,
\begin{equation}
    \label{Eq.def.S}
S_{m} \equiv  \sum_{\wt(\si_{\bm{\mu}})=m} x_{\bm{\mu}}^{2} .
\end{equation}
They define positive numbers,
\begin{equation}
\label{Ineq.Sk}
 S_{m} \geqslant 0  , \quad \text{ for } m = 1, \dots, N
\end{equation}
since they are a sum of squares. However, the values of $S_{m}$ are not arbitrary when $\rho$ varies over the set of valid density matrices. Their allowed ranges are constrained by the fundamental properties of $\rho$, including positivity, normalization, and (if applicable) purity. For example, normalization of $\rho$ implies that
\begin{equation}
\label{Eq.Trace1}
S_0 = \Tr(\rho)^2 = 1.
\end{equation}
The purity of $\rho$ can be expressed in terms of the $S_{m}$ as
\begin{equation}
\label{psipsi}   
\begin{aligned}
      \Tr \left( \rho^2 \right)
            ={}& \frac{1}{2^N} \sum_{m= 0 }^{N} \sum_{\wt(\si_{\bm{\mu}})=m} x_{\bm{\mu}}^{2}
      \\
            ={}& \frac{1}{2^N}\left(S_0+S_1+S_2 + \ldots+S_N\right) ,
\end{aligned}
\end{equation}
where we have used the orthonormality of the Pauli string basis and the expansion of $\rho$ in Eq.~\eqref{tensorrepmixed}. Since $\Tr(\rho^2)$ is bounded between $2^{-N}$ and $1$ for any density matrix, this yields the constraint
\begin{equation}
\label{Eq.Rho}
1 \leqslant S_0+S_1+S_2 + \ldots+S_N \leqslant 2^N ,
\end{equation}
where the upper bound is achieved for pure states, while the lower bound is achieved for the maximally mixed state. 
Additional constraints on the sector lengths can be derived based the time-reversed density matrix~\cite{eltschka2015monogamy}
\begin{equation}
\label{tilderho}
\tilde{\rho} \equiv \sigma_y^{\otimes N} \rho^* \sigma_y^{\otimes N}\,,
\end{equation}
where $^*$ denotes complex conjugation. The quantity
\begin{equation}
\label{Rrho}
   R_{\rho}\equiv \Tr(\rho\tilde{\rho})= \frac{1}{2^N} \sum_{\bmu} x_{\bmu} x^{\bmu}
\end{equation}
defines the overlap between the density matrix $\rho$ and its time-reversed counterpart $\tilde{\rho}$. Here, we introduced the dual coefficients $x^{\bmu}=\sum_{\boldsymbol{\nu}} g^{\bmu\boldsymbol{\nu}}x_{\boldsymbol{\nu}}$, 
where the metric tensor is defined as a product of single-qubit $g^{\mu\nu}=\textrm{diag}(1,-1,-1,-1)$. Expressed in terms of the sector lengths $S_m$, $R_\rho$ takes the form~\cite{eltschka2015monogamy}
\begin{equation}
\label{psipsipsitilde}   
R_{\rho}=\frac{1}{2^N}\left[S_0-S_1+S_2-\ldots+(-1)^NS_N\right]\,.
\end{equation}
Since $R_{\rho} \geq 0$ for all physical states, combining this with the purity constraint in Eq.~\eqref{Eq.Rho} yields the inequality
\begin{equation}
\label{Eq.Rho.Prime}
0 \leqslant S_0-S_1+S_2-\ldots+(-1)^NS_N \leqslant 2^N\, .
\end{equation}
\subsection{Constraints for pure states}
\label{purestateIneqs}
For a pure state $\rho_\psi=\ket{\psi}\bra{\psi}$, the values of $S_m=S_m(\rho_\psi)$ are further constrained by
the fact that $\Tr(\rho_\psi^2)=1$, which produces
\begin{equation}
\label{Eq.Pure.statesA}
\sum_{m=0}^N S_m = 2^{N} \, .
\end{equation}
Moreover, for any odd number of qubits $N$, the time-reversed state 
\begin{equation}
\label{Eq.Time.sta}
\ket{\tilde{\psi}}=\sigma_y^{\otimes N}\ket{\psi^*}
\end{equation} 
is orthogonal to the original pure state $\ket{\psi}$~\cite{PhysRevA.63.044301,eltschka2015monogamy}. Therefore, the overlap $R_{\rho_\psi}= |\langle \psi | \tilde{\psi} \rangle|^2=0$, and Eq.~\eqref{Eq.Rho.Prime} yields
\begin{equation}
\label{Eq.Pure.statesB}
\begin{aligned}
0 \;\leq\; \sum_{m=0}^{N} (-1)^m \, S_m \;&\leq\; 2^{N}, 
    &\quad &\text{for $N$ even,} \\[2mm]
\sum_{m=0}^{N} (-1)^m \, S_m \;&=\; 0, 
    &\quad &\text{for $N$ odd.}    
\end{aligned}
\end{equation}

We now turn to the reduced states obtained by taking partial traces of $\rho_\psi$. For any bipartition $A \cup \bar{A} = \{1, \ldots, N\}$, the purities of the reduced density matrices $\rho_A$ and $\rho_{\bar{A}}$ are equal, $\Tr (\rho_A^2)=\Tr (\rho_{\bar{A}}^2)$. This equality leads to the following set of identities~\cite{PhysRevLett.78.1600,Huber_Severini_2018} \begin{equation}
\label{symetrieS}
    \frac{1}{2^{N-k}}\sum_{m=0}^{N-k}\mbinom{N-m}{k}\,S_{m}
    = 
    \frac{1}{2^k}
    \sum_{m=0}^{k}\mbinom{N-m}{N-k}\,S_{m} ,
\end{equation}
valid for all $k = 0, \ldots, N$.
A full derivation of Eq.~\eqref{symetrieS} is provided in Appendix~\ref{App.first.Eqs}. These relations represent an alternative formulation of the MacWilliams identity, a well-known result in classical coding theory~\cite{PhysRevLett.78.1600, 681316,Huber_2018}. Importantly, they allow one to express the sector lengths $S_m(\rho_\psi)$ for $m > \lfloor N/2 \rfloor$ in terms of those with $m \leq \lfloor N/2 \rfloor$, significantly reducing the number of independent $S_m(\rho_\psi)$. 
\subsection{Shadow inequalities}
Additional constraints on the sector lengths of pure and mixed $N$-qubit states are given by the so-called \emph{quantum shadow inequalities}~\cite{681315,Rains:796376,Huber_Severini_2018}, which take the form
\begin{equation}
\label{Eq.Kravchuk}
  \Shpsi_g \equiv \frac{1}{2^N} \sum_{m=0}^N(-1)^{m} K_g(m,N)\,S_{m} \geqslant 0 \, ,
\end{equation}
for all $ g = 0, \dots , N, $
and where $K_g(m,N)$ are known as Kravchuk polynomials~\cite{krawtchouk1929,412678,681315}, defined by
\begin{equation}
\label{Eq.Krav.poly}
   K_g(m,N) \equiv \sum_{i=0}^g (-1)^i \, 3^{g-i}\mbinom{m}{i}\mbinom{N-m}{g-i}.
\end{equation}
The quantities $\Shpsi_g$ are called \emph{shadow enumerators}. In particular, for $g=0$, we recover the overlap with the time-reversed state
\begin{equation}
\label{Eq.Shadow.Rrho}
    \Shpsi_0 (\rho) = R_{\rho} \geqslant 0 . 
\end{equation}
The inverse relation of Eq.~\eqref{Eq.Kravchuk} is given by (see Appendix~\ref{App.Shor.Laflamme} for a detailed derivation)
\begin{equation}
\label{Eq.Inv.Sk}
    S_{m} =  \frac{(-1)^{m}}{2^N} \sum_{g=0}^N K_{m}(g,N) \Shpsi_g .
\end{equation}
We explain in more detail the connection between the inequalities \eqref{Eq.Kravchuk} and quantum coding theory in Appendix~\ref{App.Shor.Laflamme}. It should be noted that the inequalities~\eqref{Eq.Kravchuk} can also be derived independently of the quantum coding framework using a representation of sector lengths in the double-copy Hilbert space. For the sake of completeness, this alternative derivation is included in Appendix~\ref{App.Doublecopy}.

\subsection{Numerical range}
Taking into account the positivity condition \eqref{Ineq.Sk} and the normalization condition \eqref{Eq.Trace1}, the sector length vector $\mathbf{S}\equiv (S_1 , \dots , S_N)$ lies within a subset of $\mathbb{R}_+^N$. We define the \emph{numerical range}
of sector lengths for pure states as
\begin{equation}
    \label{defScal}
    \Scal\equiv\left\{\mathbf{S}(\rho_\psi) \Big| \, \rho_\psi=\ket{\psi}\bra{\psi}\right\}\subset \mathbb{R}_+^N\;.
\end{equation}
The main aim of our work is to characterize $\Scal$ as completely as possible by deriving inequalities that capture the monogamy of quantum correlations in pure multiqubit states. To that end, we define $\RR_1$ as the subset of vectors in $\mathbb{R}_+^N$ that satisfy the conditions specific to pure states given in Sections \ref{sec:additionalIneqs} and \ref{purestateIneqs}:
\begin{equation}
\label{Eq.Range1}
 \RR_1  = \left\{  \mathbf{S} \in\mathbb{R}_+^N\, \Big| \, \text{Eqs.~\eqref{Eq.Pure.statesA}, \eqref{Eq.Pure.statesB}, \eqref{symetrieS} hold $\forall \; k=0,\dots, N$} \right\} . 
\end{equation}
The shadow inequalities~\eqref{Eq.Kravchuk} for $g = 0, \dots , N$ impose additional constraints on $\mathbf{S}$, further refining the admissible region to
\begin{equation}
\label{Eq.Range2}
\RR_2 = \RR_1 \cap \left\{  \mathbf{S} \,\Big| \,\text{inequalities~\eqref{Eq.Kravchuk} hold $\forall \; g=0,\dots,N$}\right\}\,.
\end{equation}
By definition, we have that $\Scal\subseteq \RR_2 \subseteq \RR_1 $. In the next section, we derive new inequalities that further restrict the allowed values of the sector lengths $S_m$, thereby refining the characterization of $\Scal$.
\section{Monogamy inequalities from reduced density matrices}
\label{Sec3.Monogamy}
In this section, we derive inequalities that complement those in the previous section, starting from the Schmidt decomposition of a pure state. This will give inequalities associated with the reduced density matrices of $\ket{\psi}$ whose purities are related to the sector lengths.

\subsection{Inequalities for reduced density matrices}
Consider a pure state $\ket{\psi}$ of $N$ qubits, and a bipartition $A|\bar{A}$ that splits the system into $k$ and $N-k$ qubits.
The Schmidt decomposition of $\ket{\psi}$ with respect to this bipartition reads as
\begin{equation}
\ket{\psi}=\sum_{i=1}^{2^{\min (k,N-k)}} \sqrt{p_i}\,\ket{\phi_i}\otimes\ket{\chi_i},
\end{equation}
with $p_i\geqslant 0$, $\sum_ip_i=1$, and $\{\ket{\phi_i}\}_i$ and $\{\ket{\chi_i}\}_i$ forming orthonormal sets of states in the subsystems $A$ and $\bar{A}$, respectively. By tracing over subsystem $\bar{A}$, we obtain the reduced density matrix $\rho_A=\sum_i p_i\ket{\phi_i}\bra{\phi_i}$, and therefore
\begin{subequations}
\label{trrhoa}
\begin{align}
\label{eq:trrhoa_a}
& \Tr(\rho_A^2) = \sum_i p_i^2 = 1 - 2\sum_{i<j} p_i p_j , \\
\label{eq:trrhoa_b}
& R_{\rho_A} = \Tr(\rho_A \tilde{\rho}_A) = \sum_{i,j} p_i p_j | \braket{\phi_i | \tilde{\phi}_j} |^2.
\end{align}
\end{subequations}
where $\ket{\tilde{\phi}_j}$ is defined as in Eq.~\eqref{Eq.Time.sta}.
Using the fact that the purities of $\rho_A$ and $\rho_{\bar{A}}$ are equal for any pure bipartite state $\ket{\psi}$ and minimal when the reduced state of the smallest subsystem is maximally mixed, Eq.~\eqref{trrhoa} implies the following inequalities:
\begin{subequations}
\begin{align}
\label{Ineq.rho.A}
\frac{1}{2^{\min (k, N-k)}} \leqslant\Tr(\rho_A^2)&\leqslant  1 ,\\
0 \leqslant R_{\rho_A}&\leqslant  1 .
\label{Ineq.rho.B}
\end{align}
\end{subequations}

As noted in section \ref{purestateIneqs}, for any odd $k$, a $k$-qubit state $\ket{\phi}$ is orthogonal to its time-reversed state $\ket{\tilde{\phi}}$. Thus, for odd $k$, the expression for $R_{\rho_A}$ simplifies to $2\sum_{i<j}p_ip_j\vert\braket{\phi_i\vert\tilde{\phi}_j}\vert^2$. Combining with Eq.~\eqref{eq:trrhoa_a}, we obtain
\begin{equation}
\label{rrhoatrrhoa}
1- \Tr(\rho_A^2) - R_{\rho_A} =2\sum_{i<j}p_ip_j\left(1-\vert\braket{\phi_i\vert\tilde{\phi}_j}\vert^2\right) \geqslant 0 .
\end{equation}
Summing Eqs.~\eqref{Ineq.rho.A} and \eqref{Ineq.rho.B}, we obtain the following proposition:
\begin{prop}
\label{Prop.1}
For a general pure state of $N$ qubits
    \begin{equation}
\label{Ineq.rhoA.RA}
2^{-\min(k,N-k)}
 \leqslant \mathrm{Tr}(\rho_A^2) + R_{\rho_A} \leqslant \frac{3+(-1)^k}{2}\, .
\end{equation}
\end{prop}
The upper bound of the latter inequality, for odd $k$, follows from Eq.~\eqref{rrhoatrrhoa}.
\subsection{Sector length inequalities}
The quantities $\Tr(\rho_A^2)$ and $R_{\rho_A}$ in Proposition~\ref{Prop.1} can be expressed in terms of the sector lengths $S_{m}(\rho_A)$. Indeed, using Eqs.~\eqref{psipsi} and \eqref{psipsipsitilde} for $\rho_A$ (a mixed $k$-qubit state), we get
\begin{equation}
\label{trArA}
\begin{aligned}
 & \Tr \left( \rho_A^2 \right) = \frac{1}{2^k}\sum_{m=0}^k
  S_{m}(\rho_A) ,
  \\
& R_{\rho_A}= \frac{1}{2^k}\sum_{m=0}^k(-1)^{m}S_{m}(\rho_A)\,.    
\end{aligned}
\end{equation}
In order to obtain additional inequalities for the sector lengths of the pure state $\rho_\psi=\ket{\psi}\bra{\psi}$ of which $\rho_A$ is a reduction, we consider sums over all sets $A$ with $k$ elements. For fixed $k$ and $m \leqslant k$, the sector lengths $S_{m}(\rho_A)$ of the reduced states and the sector lengths $S_{m}(\rho_\psi)$ of the pure state are related by the identity
\begin{equation}
\label{sumaSi}
    \sum_{|A|=k} S_{m} \left(\rho_A \right)=\mbinom{N-m}{k-m}S_{m}(\rho_\psi)\,\quad \forall\,m \leqslant k \, ,
\end{equation}
where $|A|$ denotes the cardinality of $A$.
A proof of \eqref{sumaSi} is given in Appendix~\ref{App.first.Eqs}. By summing~\eqref{trArA} over all subsets of length $k$ and using \eqref{sumaSi}, we obtain
\begin{align}
\label{rhoSSgen}
& \sum_{|A|=k} \Tr(\rho_A^2)=\frac{1}{2^k}\sum_{m=0}^k \mbinom{N-m}{k-m}S_{m}(\rho_\psi) ,
   \\
& \sum_{|A|=k}  R_{\rho_A}=\frac{1}{2^k}\sum_{m=0}^k (-1)^{m}\mbinom{N-m}{k-m}S_{m}(\rho_\psi) .
\label{RSSgen}
\end{align}
Summing Eqs.~\eqref{Ineq.rho.A}, \eqref{Ineq.rho.B} and the result in Proposition~\ref{Prop.1} in the same way, we get
\begin{equation}
\label{Ineq.Schmidt.Pur}
\begin{aligned}
\frac{1}{2^{\min (k,N-k)}}\mbinom{N}{k}  & \leqslant  \sum_{|A|=k} \Tr(\rho_A^2)\leqslant \mbinom{N}{k} ,
\\
 0  & \leqslant \sum_{|A|=k}  R_{\rho_A}\leqslant 
\mbinom{N}{k} ,
\\
 \frac{\mbinom{N}{k} }{2^{\min (k,N-k)}} & \leqslant \sum_{|A|=k} \left[
\Tr(\rho_A^2)+R_{\rho_A}  \right]\leqslant \frac{3+(-1)^k}{2} \mbinom{N}{k}.
\end{aligned}
\end{equation}
This leads to new constraints on the sector lengths $S_m=S_m(\rho_\psi)$ that must be satisfied by any pure multiqubit state:
\begin{align}
\label{Eq.Ine.R3.Eq1}
& \frac{2^k}{2^{\min(k,N-k)}} \mbinom{N}{k}  \leqslant 
\sum_{m=0}^k \mbinom{N-m}{k-m} S_{m}  \leqslant 2^k \mbinom{N}{k} ,
\\
\label{Eq.Ine.R3.Eq2}
& 0\leqslant \sum_{m=0}^k (-1)^{m}\mbinom{N-m}{k-m}S_{m} \leqslant 2^k \mbinom{N}{k} ,
\\
\label{Eq.Ine.R3.Eq3}
&\sum_{m=0}^{\lfloor\frac{k-1}{2}\rfloor}\mbinom{N-2m}{k-2m}S_{2m} \leqslant
2^{k-2} \left( 3 + (-1)^k \right) \mbinom{N}{k}. 
\end{align}
Equations \eqref{Eq.Ine.R3.Eq1}--\eqref{Eq.Ine.R3.Eq3} define a new region in the $\mathbf{S}$ space that further restricts the admissible values of the sector lengths. We denote this region by
\begin{equation}
\label{Eq.Range3}
\RR_3 = \RR_1 \cap \left\{  \mathbf{S} \, \Big| \, \text{Eqs.~\eqref{Eq.Ine.R3.Eq1}--\eqref{Eq.Ine.R3.Eq3} hold $\forall\;k=1, \dots, N$} \right\} .
\end{equation}
We note that the upper bound in Eq.~\eqref{Eq.Ine.R3.Eq1} and the lower bound in Eq.~\eqref{Eq.Ine.R3.Eq2} were previously established in Proposition 7 and Eq.~(32) of Ref.~\cite{Huber_Severini_2018}.

\subsection{Summary of our results}
We summarize our main result in the following theorem.
\begin{theorem}
    \label{Theo.RR.region}
Let 
\begin{equation}
\label{defR}
\RR = \RR_2\cap \RR_3 \, ,
\end{equation}
where $\RR_2$ and $\RR_3$ are the subsets of $\mathbb{R}_+^N$ defined in Eqs.~\eqref{Eq.Range2} and~\eqref{Eq.Range3}, respectively. Then, $\RR$ defines a polytope and the set $\Scal$ of achievable sector length vectors for pure states is contained within this polytope, that is,
\begin{equation}
\Scal \subset \RR.
\end{equation}
\end{theorem}

Although numerical optimization of $S_m$ on the simplex $\RR$ is relatively straightforward, finding the extremal values of $S_m$ within the set $\mathcal{S}$ involves optimizing a quartic polynomial in $2^{N+1}$ variables (the real and imaginary parts of the components of $\ket{\psi}$). This is computationally feasible only for a small number of qubits.

We performed a numerical optimization to determine the extremal values of the $S_m$ and the corresponding pure states that achieve these optima, for $m=1,2,\ldots, N$ on systems of $N=1,\dots ,8$ qubits. The results are summarized in Table~\ref{tab.3}. The states highlighted in gray can be shown analytically to saturate the bounds defining the set $\RR$. Since $\Scal \subset \RR$, this confirms that these states are indeed the optimal pure states within $\Scal$. Some of these optimal states were previously known, with references provided in Table~\ref{tab.3}. 

The following list presents results that, to our knowledge, have not been reported before:
\begin{itemize}
    \item[--] For $N=4$, $\Scal$ is completely characterized and coincides with the polytope $\RR$.
    \item[--] For $N=5$, the boundaries of $\Scal$ are fully characterized, and strong numerical evidence suggests that $\Scal$ coincides with the polytope $\RR$.
    \item[--] For $N = 5$ and $6$, it is known that absolutely maximally entangled (AME) states satisfy $S_2 = 0$. Furthermore, we show that the specific AME states given in Eqs.~\eqref{Eq.AME52.def} and \eqref{Eq.AME62.def} also achieve the global maximum of $S_4$. It should be noted that there are other six-qubit AME states that do not achieve this maximum.
    \item[--] For $N=5$, the global minimum of $S_4$ is reached by a product state.
    \item[--] For $N=6$, the symmetric state given in Eq.~\eqref{pyramidN6} maximizes $S_5$.
    \item[--] We find the minimum number of qubits $N_0$ for which Theorem~\ref{Conj.Wyderka.GuhneTHM} holds, which concerns the maximum value of $S_m$, for $m=4$ and $5$.
    \item[--] We find states with $S_1=S_2=0$ for any $N\geq 5$ in Sec.~\ref{Subsec.1.2.body}.
\end{itemize}

Furthermore, we derive another proof that the maximum value of $S_N$ is equal to $2^{N-1}+1$ for $N$ even. This result was conjectured in Ref.~\cite{PhysRevLett.108.240501} and later proved in Ref.~\cite{Eltschka2020}. Our alternative proof is presented in Subsection~\ref{Subsec.Upper.SN} and is based entirely on our inequalities.
\section{Monogamy relations for small numbers of qubits}
\label{Sec4.Num.Range}
In this section, we apply Theorem~\ref{Theo.RR.region} to determine the numerical range $\Scal$ of the sector lengths of pure multiqubit states~\eqref{defScal}. Our analysis focuses on systems between two and six qubits. For $N\leq 5$, we show that $\Scal$ forms a polytope. In these cases, we also identify the extremal quantum states corresponding to the vertices and edges of the respective polytopes. However, the case $N = 6$ presents a significantly more complicated scenario, and we are unable to verify that $\Scal$ retains a polytope structure. Table~\ref{tab.1} summarizes optimal quantum states associated with the extremal points of the set $\RR$ for all $N\leq 6$. These states account for all upper bounds reported in Table~\ref{tab.3}, and also for all lower bounds, with the exception of the bound $S_4 \geqslant 5$ for $N = 6$. To express these results, we make use of the generalized Greenberger-Horne-Zeilinger (GHZ) state, defined as
\begin{equation}
\label{Eq.Gen.GHZ}
    \ket{\mathrm{GHZ}(N,\phi)} \equiv \cos \big( \tfrac{\phi}{2}\big) \ket{0}^{\otimes N} 
    + \sin \big( \tfrac{\phi}{2} \big) \ket{1}^{\otimes N}
\end{equation}
with $\phi\in [0,\pi]$. We denote the standard $N$-qubit GHZ state by $\ket{\mathrm{GHZ}(N) } \equiv \ket{\mathrm{GHZ}(N, \frac{\pi}{2})}$.
\subsection{\texorpdfstring{Two qubits}{Lg}}
In this case, $\RR_1\subset \mathbb{R}_+^2$ is defined by the following nontrivial (in)equalities
\begin{equation}
\label{conditionsN2}
    S_1 + S_2 = 3 ,
\qquad
0 \leqslant 1- S_1 + S_2 \leqslant 4.
\end{equation}
These conditions imply that $S_1\in [0,2]$, $S_2\in [1,3]$, and that $S_1$ and $S_2$ are linearly related. The sets $\RR_2$ and $\RR_3$ coincide with $\RR_1$, so all three are fully characterized by the constraints \eqref{conditionsN2}. For the generalized GHZ state $\ket{\mathrm{GHZ}(2,\phi)}$, the sector lengths are given by $S_1=2 \cos^2 \phi$ and $S_2=3-2 \cos^2 \phi$. When $\phi$ varies, these expressions span the set of allowed values $(S_1,S_2)$, with the extremal values being achieved by product states and by the $\ket{\mathrm{GHZ}(2)}$ state. This shows that each point in $\RR$ corresponds to a pure state. It follows that $\Scal=\RR$, and that the entire set $\Scal$ can be parametrized by the angle $\phi$. Note that the set of achievable sector lengths has also been characterized for mixed two-qubit states in Ref.~\cite{wyderka2020characterizing}. By relaxing the purity constraint, this set becomes significantly larger and more complex to characterize.

\subsection{\texorpdfstring{Three qubits}{Lg}}
For three qubits, the set $\RR_1\subset\mathbb{R}_+^3$ is defined by the constraints
\begin{equation}
\begin{aligned}
S_2=3 ,\qquad S_1 + S_3 ={}& 4 .
\end{aligned}
\end{equation}
The smaller sets $\RR_2$ and $\RR_3$ coincide, and differ from $\RR_1$ by the additional inequality
\begin{equation}
0 \leqslant S_1 \leqslant 3.
\end{equation}
Thus, it follows that $\RR = \RR_2 = \RR_3$ is delimited by the intervals $S_1 \in [0,3]$ and $S_3 \in [1,4]$, with $S_2=3$. As in the case of two qubits, the family of states $\ket{\mathrm{GHZ}(3,\phi)}$ leads to $S_1 = 3\cos^2 \phi$, which spans the entire interval $\RR$. Therefore, $\RR = \Scal$, with the extremal values attained by product states and the $\ket{\mathrm{GHZ}(3)}$ state. As for $N=2$, the set of achievable sector lengths for mixed three-qubit states has also been fully characterized in Ref.~\cite{wyderka2020characterizing}, which contains the particular values for pure states described above.

\begin{figure}[t]
    \centering
\includegraphics[width=.38\textwidth]{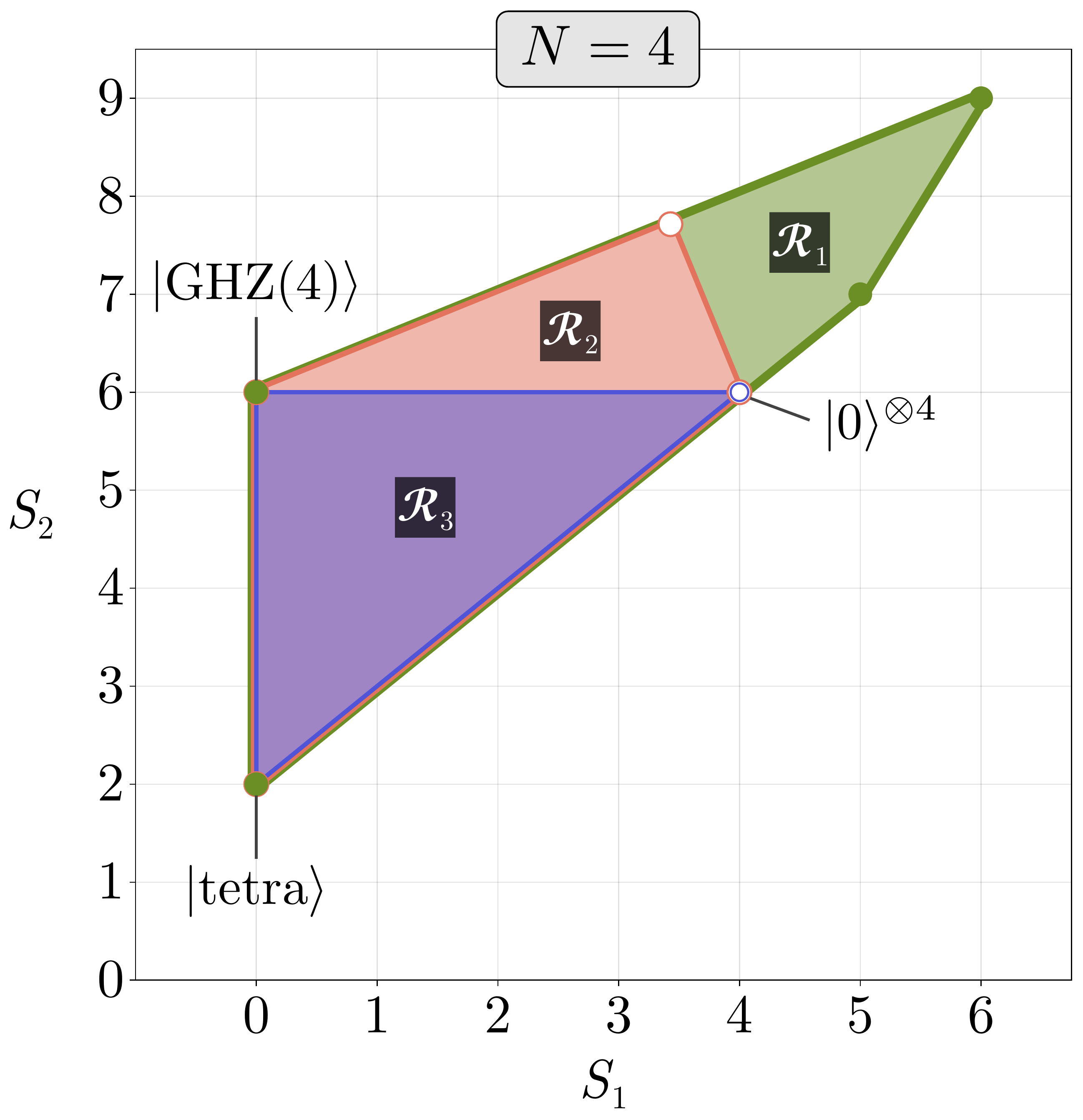}\\[10pt]
\includegraphics[width=.38\textwidth]{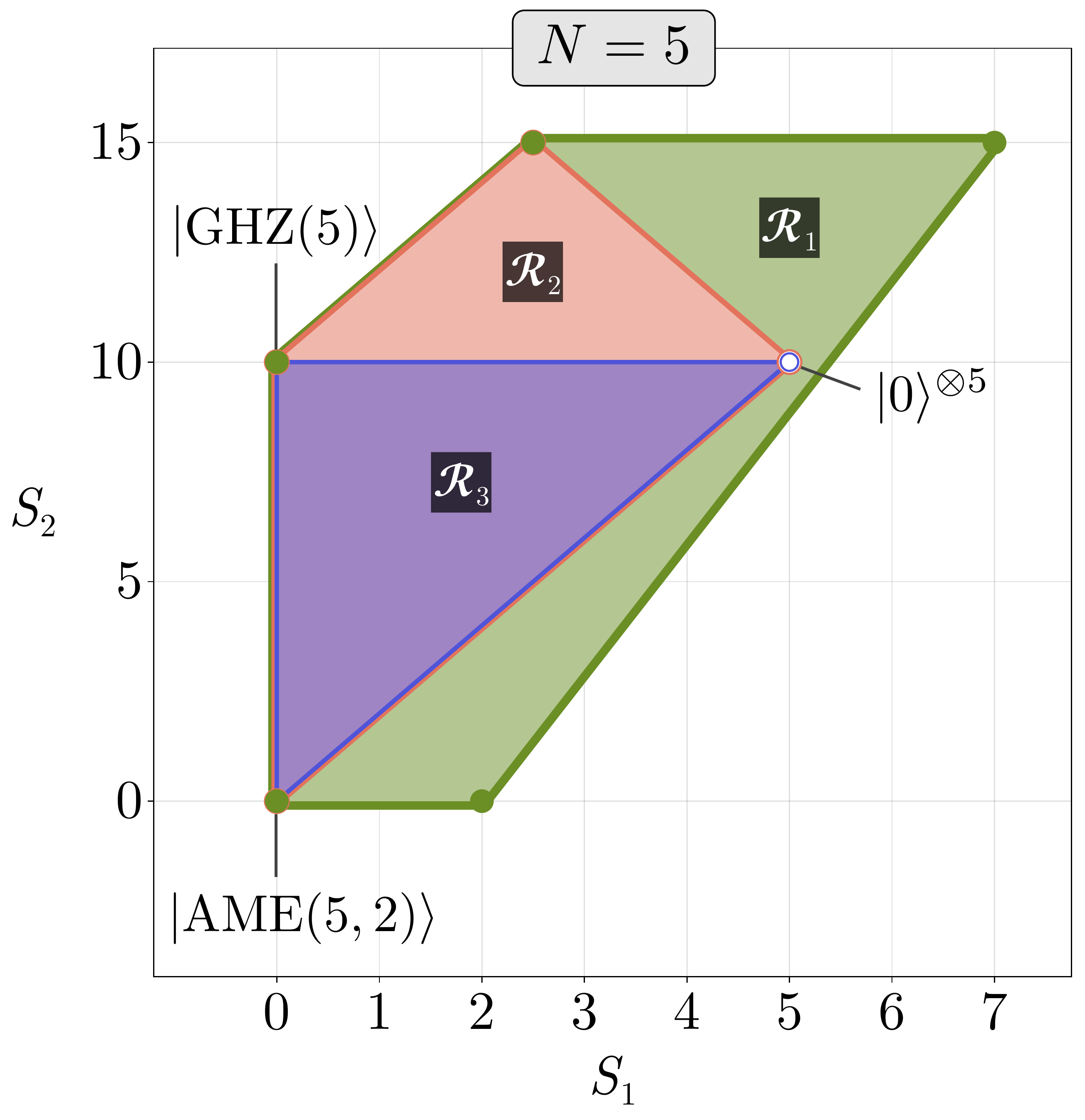}
    \caption{
    Projection of the sets $\RR_3 \subset \RR_2 \subset \RR_1 $ onto the $(S_1 ,S_2)$ plane for $N=4$ (top) and $N=5$ qubits (bottom), colored in purple, red and green, respectively. For these particular numbers of qubits, $\RR = \RR_3$ and every point in $\RR_3$ corresponds to a physically realizable quantum state.}
    \label{fig1}
\end{figure}
\subsection{\texorpdfstring{Four qubits}{Lg}}
For four qubits, $\RR_1\subset\mathbb{R}_+^4$ is specified by the equalities
\begin{equation}
\label{sl4sl5}
    S_4 = 3-2S_1+S_2 , \qquad 
    S_3 = 12+S_1-2S_2 ,
\end{equation}
and the inequality
\begin{equation}
\label{N4ineq}
    2 \leqslant S_2 - S_1 \leqslant 6.
\end{equation}
In Fig.~\ref{fig1}, we plot the projection of $\RR_1$ onto the sector length subspace $(S_1 ,S_2)$ (shown as the largest green region). While the set $\RR_2$ (shown in red) is reduced by the additional inequality $3 S_1+S_2 \leqslant 18$, the set $\RR_3$ (shown in purple)  is reduced by an even stricter inequality $S_2 \leqslant 6$, such that $\RR = \RR_3 \subset \RR_2 \subset \RR_1$. The intervals delimiting the set $\RR_3$ are $S_1 \in [0, 4]$, $S_2 \in [2, 6]$, $S_3 \in [0, 8]$, and $S_4 \in [1, 9]$, with the limits reached by the states indicated in Table~\ref{tab.3}.  It should be noted that this case has also been studied in Refs.~\cite{10.1063/1.3511477,10.1063/1.2435088}: in terms of the bipartite tangles $\tau_1$ and $\tau_2$ introduced there, we have $S_1=4(1-\tau_1)$ and $S_2=6+12(\tau_1-\tau_2)$, and the inequalities $\tau_1\leq \tau_2\leq \frac43\tau_1$ obtained in ~\cite{10.1063/1.3511477,10.1063/1.2435088} correspond to $S_2\leq 6$ and $2\leq S_2-S_1$, respectively.

The set $\RR = \RR_3 $, which is entirely characterized by the conditions 
\begin{equation}
\label{ineqRn4}
 0\leqslant S_1\leqslant  S_2-2\leqslant 4,
\end{equation}
corresponds to the purple triangle in the top panel of Fig.~\ref{fig1}. A comparable plot can be found in Ref.~\cite{Wyderka_thesis}, where examples of pure states corresponding to the boundary of the set $\RR$ are given. Here, we find a family of states that explicitly realizes all points in $\RR$. It is given by
\begin{equation}
\ket{\psi_4(\theta,\phi)}= \cos\left(\tfrac{\theta}{2}\right)\ket{\textrm{GHZ}(4,\phi)}+ i \sin\left(\tfrac{\theta}{2}\right) \ket{D_4^{(2)}}
\label{familypsi4}
\end{equation}
with $\theta\in[0,\frac{\pi}{2}[$, $\phi\in[0,2\pi]$, and the symmetric Dicke state $\ket{D_4^{(2)}}$ as defined in Eq.~\eqref{Eq.Dicke}. 
For the family \eqref{familypsi4}, the sector lengths are 
\begin{align}
\label{sl1N4}
   S_1&=(1+\cos\theta)^2\cos^2\phi,\\[4pt]
   S_2&=2 \left[3 -(1+\sin\phi)\sin^2\theta\right].
\label{sl2N4}
\end{align}
To show that Eqs.~\eqref{sl1N4} and \eqref{sl2N4} parametrize the whole set $\RR$, we let $u=\cos\theta$ and $v=\sin\phi$, which yields
\begin{equation}\label{sl1u2v}
\begin{aligned}
& S_1=(1+u)^2(1-v^2)\equiv x,\\[4pt]
& 3-S_2/2=(1+v)(1-u^2)\equiv y.
\end{aligned}
\end{equation}
Equation \eqref{ineqRn4} yields $0\leqslant y\leqslant 2-x/2\leqslant 2$.
We thus want to show that for any $x,y\geqslant 0$ such that $x+2y\leqslant 4$ one can find $u\in [0,1]$ and $v\in [-1,1]$ such that $x=(1+u)^2(1-v^2)$ and $y=(1+v)(1-u^2)$. If $y=0$ we take $u=1$ and $v\in [-1,1]$. Otherwise, we get in particular the condition
$v=1-\frac{x(1-u)}{y(1+u)}$, and $u$ must be a root of the order-2 polynomial $P(u)=(x+2y)u^2-2 u x+x-2y+y^2$. This polynomial reaches its minimum at $u_0=x/(x+2y)\in [0,1]$. The condition $x+2y\leqslant 4$ implies that $P(u_0)=y^2(x+2y-4)/(x+2y)\leqslant0$, which, together with $P(1)=y^2\geqslant 0$, ensures that there is a real root $u\in[u_0,1]$. To show that the corresponding $v=1-\frac{x(1-u)}{y(1+u)}$ is in $[-1,1]$ one needs to check that $x(1-u)\leqslant 2y(1+u)$, that is, $u\geqslant u_0-2y/(x + 2 y)$. This is true since $u\geqslant u_0$ and $x,y\geqslant 0$.
Thus there exists a solution $(u,v)$ inverting Eq.~\eqref{sl1u2v}.
This shows that $\Scal=\RR$ is given by \eqref{ineqRn4} and parametrized by the two angles $\theta$ and $\phi$ according to \eqref{sl1N4} and \eqref{sl2N4}. In particular, the maximum of $S_1$ is achieved by the state $\ket{\psi_4(0,0)}= \ket{0}^{\otimes 4}$. The minimum of $S_1$ can be reached, for instance, by $\ket{\psi_4(0,\pi/2)}= \ket{\mathrm{GHZ}(4)}$. Lastly, $\min S_2=2$ is achieved by the state $\ket{\psi_4(\pi/2,\pi/2)}= \ket{\mathrm{tetra}}$ [see Eq.~\eqref{tetrapsi} for more details].
\subsection{\texorpdfstring{Five qubits}{Lg}}
In this case, $\RR_1\subset\mathbb{R}_+^5$ is defined by the equalities
\begin{equation}
\begin{aligned}
S_3 ={}& 10+2 S_1-S_2 , 
\\
S_4 ={}& 15-S_2 ,
\\
S_5 ={}& 6-3 S_1+S_2 .
\end{aligned}
\end{equation}
and the inequalities
\begin{equation}
\begin{aligned}
   S_2 \leqslant 15 ,\quad 
   3S_1 -6 \leqslant S_2 \leqslant 2S_1 +10.
\end{aligned}
\end{equation}
$\RR_2$ is constrained by the two additional inequalities
\begin{equation}
    2S_1 \leqslant S_2 , \qquad 2S_1+S_2 \leqslant 20.
\end{equation}
Finally, $\RR_3$ is characterized by the above inequalities, together with the additional condition $S_1 \leqslant 10$. In Fig.~\ref{fig1} we plot the projection onto the $(S_1 ,S_2)$ plane of the sets $\RR_3 \subset \RR_2 \subset \RR_1$. The intervals of the sector lengths $S_m$ are $S_1\in [0,5]$,  $S_2\in [0,10]$, $S_3\in [0,10]$, $S_4\in [5,15]$ and $S_5\in [1,16]$. The endpoints of these intervals are reached by the examples given in Table~\ref{tab.3}. 

The region $\RR$ is now characterized by the conditions 
\begin{equation}
\label{ineqRn5}
 0\leqslant S_1\leqslant 5,\qquad 2S_1\leqslant S_2\leqslant 10, 
\end{equation}
corresponding to the purple triangle in Fig.~\ref{fig1} (bottom panel). 
As shown in the previous subsections, for $N\leqslant 4$ one can realize all points in $\RR$ by simply considering a family of pure states made of superpositions of extremal points. This is no longer the case for $N=5$, and finding a state realizing an arbitrary point of the domain $\RR$ is far from trivial. 

Let us consider the left boundary of $\RR$, that is, points for which $S_1=0$ and $0\leqslant S_2\leqslant 10$. A superposition of the $\ket{\mathrm{GHZ}(5)}$ and $\ket{\mathrm{AME}(5,2)}$ states does not enable to reach all values of $S_2$ for $S_1=0$. However, we can exhibit a family of states which realizes all points on this boundary. Setting $z=t x - (t + 1) y$, we found that states of the form 
\begin{multline}
    \ket{\psi}=
(0,-t z,0,x,y,0,t z,0,z,0,z,0,0,y,0,z,\\
-z,0,-y,0,0,z,0,z,0,t z,0,y,-x,0,t z,0),
\label{purestateN5}
\end{multline}
in the computational basis, depending on three parameters $x,y,t$, are such that $S_1=0$. The normalization condition for $\ket{\psi}$ can be rewritten
\begin{multline}
        2 \left(2 t^4+3 t^2+1\right) x^2-4 t \left(2 t^3+2 t^2+3 t+3\right) x y
       \\
        +2\left(2 t^4+4 t^3+5 t^2+6 t+5\right)y^2=1,
\label{constraintS2N5}
\end{multline}
while $S_2$ is a polynomial in $(x,y,t)$ given by
\begin{align}
  &\frac{1}{8}S_2=-4 t (t+1)^2 \left(8 t^5+8 t^4+t+17\right) xy^3\nonumber\\
   &+2 \left(24 t^8+48 t^7+24 t^6+15 t^4+70 t^3+54 t^2-2 t-1\right) x^2 y^2\nonumber\\
  &\!\!\!\!+\left(8t^8+32 t^7+48 t^6+32 t^5+5 t^4+4 t^3+30 t^2+36 t+21\right) y^4\nonumber\\
 &+\left(8 t^8+13 t^4-2 t^2+5\right) x^4\nonumber\\
  &-4 t \left(8 t^7+8t^6+9 t^3+17 t^2-t-1\right) x^3 y.
\label{exampleS2N5}
\end{align}
Analytical minimization and maximization of \eqref{exampleS2N5} under the constraint \eqref{constraintS2N5} shows that the bounds $S_2=0$ and $10$ are reached for $(x,y,t)=(\frac14, -\frac14,-1)$ and $(-\frac{1}{\sqrt{2}},0,0)$, respectively. By continuity, the family of states \eqref{purestateN5} covers the whole left boundary of the domain $\RR$. Now consider the oblique boundary of $\RR$ on the right, characterized by $S_2=2S_1$ with $S_1\in [0,5]$. We present two families of states that, taken together, cover this boundary. First, consider the family of states
\begin{multline}
    |\psi(\eta)\rangle = \frac{\cos\eta}{2}\, (|00011\rangle+|00110\rangle-|10001\rangle+|10100\rangle)\qquad\\
     + \frac{\sin\eta}{2}\,(|01000\rangle-|01101\rangle+|11010\rangle+|11111\rangle).
\end{multline}
These states have sector lengths
\begin{equation}
    S_2= 2 S_1 = 2 \cos^2(2\eta)
\end{equation}
and thus cover the segment of the right boundary of $\RR$ with $S_1 \in [0,1]$ for $\eta \in [0, \pi/4]$. Second, consider the family of states
\begin{equation}\label{superposition_psi5}
    |\phi(\eta)\rangle = \mathcal{N} \Big[\cos(\eta/2)\, |\psi_5\rangle + \frac{\sin(\eta/2)}{\sqrt{2}}\, |00000\rangle\Big],
\end{equation}
where $|\psi_5\rangle$ is given in Eq.~\eqref{psi_5} and $\mathcal{N} =2\,(\sin \eta+\cos\eta+3)^{-1/2}$ is a normalization constant. These states exhibit sector lengths given by
\begin{equation}
    S_2= 2 S_1 = \frac{28 \sin \eta-6 \sin (2 \eta)-4 \cos \eta-\cos (2 \eta)+37}{(\sin \eta+\cos \eta+3)^2}.
\end{equation}
In this case, $S_1$ increases monotonically from $1$ at $\eta = 0$ to $5$ at $\eta = \pi$, covering the right boundary of $\RR$ with $S_1 \in [1,5]$. Together, these results show that there are states that realize all the points on the right-hand boundary. Finally, the top boundary of $\RR$ is covered by the generalized GHZ state \eqref{Eq.Gen.GHZ} as $\phi$ varies within $[0, \pi/2]$, yielding sector lengths $S_1=5\cos^2\phi$ and $S_2=10$. By forming superpositions of states located at the entire boundary, we find numerically that we can obtain any combination of sector lengths within $\RR$. 

This leads us to the following conjecture:
\begin{conjecture}
\label{poltytopeN5}
For a 5-qubit system, the set of sector lengths attainable by pure states $\Scal$ is exactly the polytope $\RR$.
\end{conjecture}
\begin{figure*}[t!]
    \centering   
\includegraphics[width=.95\textwidth]{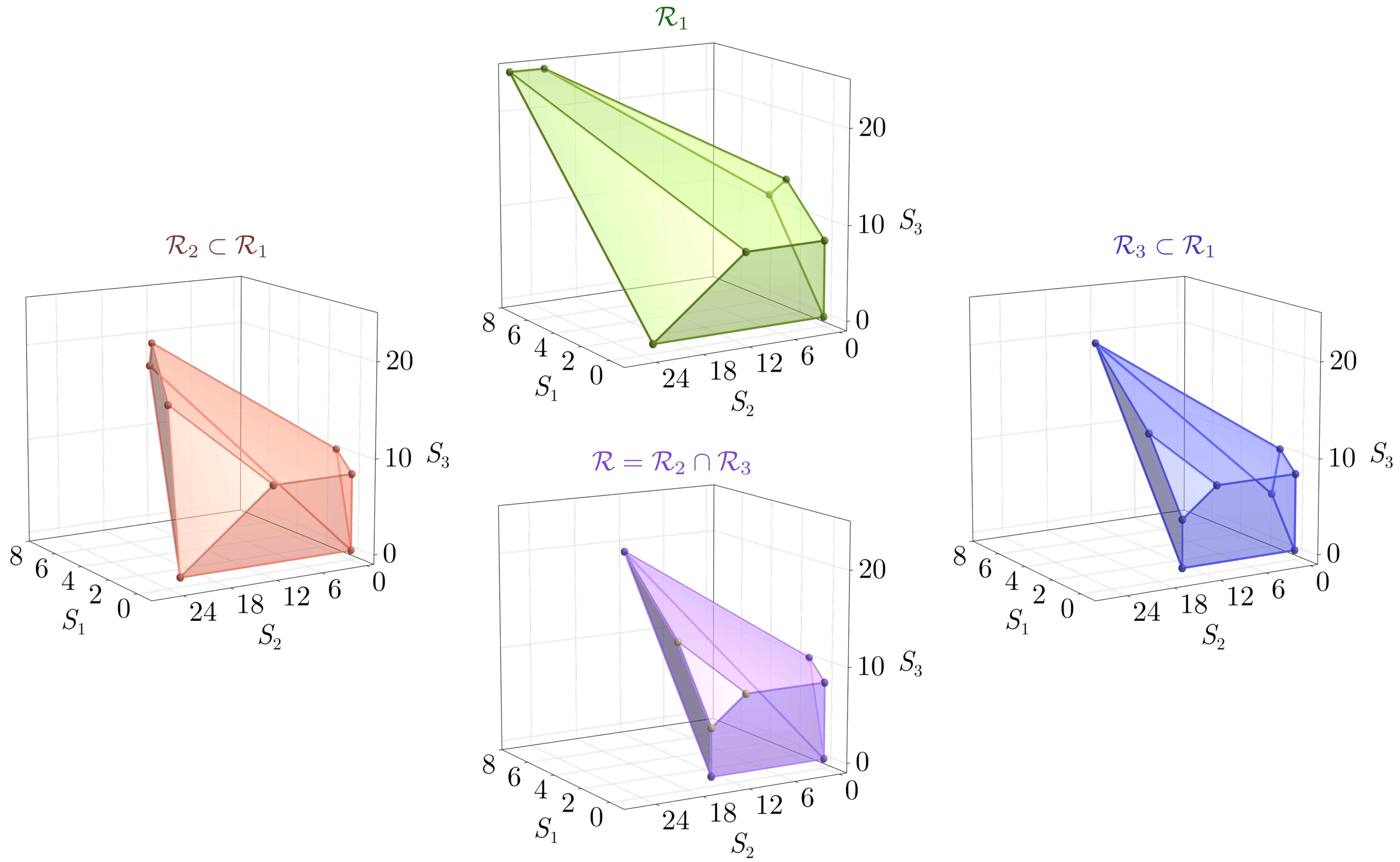}
    \caption{Projection of the sets $\RR_1$, $\RR_2 \subset \RR_1$, $\RR_3 \subset \RR_1$, and $\RR=\RR_2\cap \RR_3$ onto the $(S_1 ,S_2, S_3)$ subspace for $N=6$. Note that $\RR_3$ is not contained in $\RR_2$ and vice versa (see the differences in the faces of the polyhedra in the planes $S_2=0$ and $S_1=0$).}
    \label{fig2}
\end{figure*}

\begin{figure}[t]
    \centering   
\includegraphics[width=.45\textwidth]{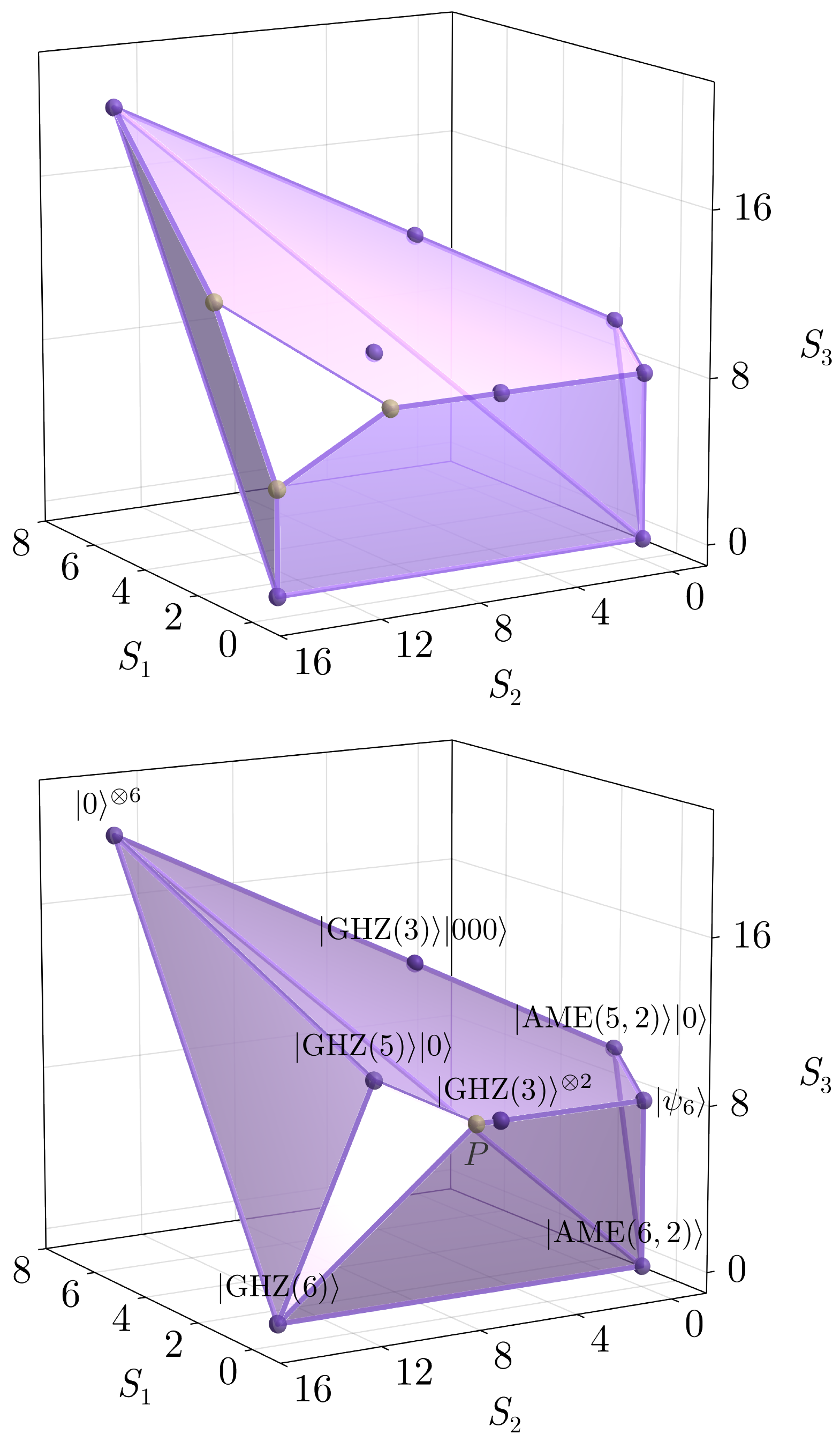}
    \caption{Top: the polytope $\RR=\RR_2 \cap \RR_3$ for $N=6$ in the $(S_1,S_2,S_3)$ space. Bottom: smaller polytope obtained by including additional linear constraints given in Eq.~\eqref{reducedpolytopeN6}, derived by numerical optimization. States corresponding to vertices or points along the edges are indicated. The vertex $P=(0,7,8)$, shown in grey, is one for which we have not identified any state realizing these sector lengths.}
    \label{fig3}
\end{figure}

\subsection{\texorpdfstring{Six qubits}{Lg}}
For $N=6$, we have three free variables $S_1$, $S_2$, and $S_3$, and
\begin{equation}
\label{equalitiesN6}
\begin{aligned}
    S_4 & = 45 + 10 S_1-2 S_2-3 S_3 ,
    \\
    S_5 & = 3 S_3 -9 S_1 ,
    \\
    S_6 & = 18 -2 S_1+S_2-S_3 .
\end{aligned}    
\end{equation}
$\RR_1$ is the subset of $\mathbb{R}_+^6$ defined by the equalities \eqref{equalitiesN6} and the inequalities:
\begin{equation}
-8 \leqslant 2S_1 - S_3 \leqslant 0.
\end{equation}
$\RR_2$ is constrained by $\RR_1$ together with the following additional inequalities
\begin{equation}
    30 S_1\leq 8 S_2+3 S_3 , \quad  10 S_1+16 S_2+3 S_3\leq 360.
\end{equation}
Similarly, $\RR_3$  is constrained by $\RR_1$ together with 
\begin{equation}
 10 S_1+S_3 -20 \leqslant 4 S_2 \leqslant 60 , \qquad 10 S_1\leq 3 S_3 .
\end{equation}
Interestingly, in this case, $\RR_2$ is not contained in $\RR_3$, as we can observe in Fig.~\ref{fig2}. Thus, $N=6$ is the smallest number of qubits for which $\RR = \RR_2 \cap \RR_3 \neq \RR_3$ and $\RR_2$. $\RR$ has the following ranges in the sector lengths: $S_1\in [0,6]$,  $S_2\in [0,15]$, $S_3\in [0,20]$, $S_4\in [0,45]$, $S_5\in [0,24]$ and $S_6\in [1,33]$. Our numerical results show that $\Scal$ is strictly contained in $\RR$, further restricted by $S_4 \geqslant 5$ (see Table~\ref{tab.3}). By integrating this constraint into the region $\RR$, along with additional linear constraints derived from numerical optimization, we obtain a reduced polytope in $\mathbb{R}_+^6$ (see the lower panel of Fig.~\ref{fig3}), whose non trivial faces are defined by the following inequalities:
\begin{equation}
\label{reducedpolytopeN6}
    \begin{aligned}
    10 S_1 - 3 S_3 &\leq 0, \\
    30 S_1 -8 S_2 -3 S_3 &\leq 0, \\
    - 10 S_1 + 4 S_2 + 3 S_3 &\leq 60, \\
    - 5 S_1 + S_2 + S_3 &\leq 15, \\
    - 2 S_1 + S_3 &\leq 8.
\end{aligned}
\end{equation}
Although this reduced polytope contains $\mathcal{S}$, the exact shape of $\mathcal{S}$ for $N=6$ remains unknown. The question of whether it is even a polytope remains open.
\subsection{\texorpdfstring{Seven and eight qubits}{Lg}}
We expect the set $\Scal$ to exhibit a complex shape for $N=7$ and $8$, similar to the case $N=6$.  Indeed, we find that, while several optimal states for the sector lengths are achieved by $\RR$ (states in a grey cell in  Table~\ref{tab.3}), numerical searches reveal other states that produce tighter bounds for certain sector lengths. Therefore, additional conditions are needed to fully characterize $\Scal$. 
\begin{table}[t!]
    \centering
    \renewcommand{\arraystretch}{1.5}
\mbox{
    \begin{tabular}{|c|c|@{\hspace{3pt}}c@{\hspace{3pt}}|}
        \hline
        \multirow{2}{*}{\textbf{$N$}} & \multirow{2}{*}{State} & $(S_1)$ or 
        \\
        & & $(S_1,S_2)$
        \\ \hline\hline
        \multirow{2}{*}{$\;2\;$} 
                    & $|\mathrm{GHZ}(2)\rangle$   & $(0)$ 
        \\ \cline{2-3}
                    & $|0\rangle^{\otimes 2}$ & $(2)$ 
        \\ \hline\hline
        \multirow{2}{*}{$\;3\;$} 
                    & $|\mathrm{GHZ}(3)\rangle$   & $(0)$ 
        \\ \cline{2-3}
                    & $|0\rangle^{\otimes 3}$ & $(3)$ 
        \\ \hline\hline
        \multirow{3}{*}{$\;4\;$} & $|\mathrm{tetra}\rangle$ & $(0, 2)$ \\ \cline{2-3}
                               & $|\mathrm{GHZ}(4)\rangle$   & $(0, 6)$ \\ \cline{2-3}
                               & $|0\rangle^{\otimes 4}$ & $(4, 6)$ 
        \\ \hline\hline
        \multirow{3}{*}{$\;5\;$} & $|\mathrm{AME}(5,2)\rangle$ & $(0, 0)$ \\ \cline{2-3}
                               & $|\mathrm{GHZ}(5)\rangle$   & $(0, 10)$ \\ \cline{2-3}
                               & $|0\rangle^{\otimes 5}$ & $(5, 10)$ 
        \\ \hline
    \end{tabular}
    
    \begin{tabular}{|c|c|c|}
        \hline
        \textbf{$N$} & State & $(S_1,S_2,S_3)$ 
        \\ \hline\hline
        \multirow{8}{*}{$\;6\;$} & $|\mathrm{AME}(6,2)\rangle$ & $(0, 0, 0)$ \\ \cline{2-3}
                               & $|\psi_6\rangle$        & $(0, 0, 8)$ \\ \cline{2-3}
                               & $|\mathrm{GHZ}(6)\rangle$  & $(0, 15, 0)$ \\ \cline{2-3}
                               & $|0\rangle^{\otimes 6}$ & $(6, 15, 20)$ \\ \cline{2-3}
                               & $|\mathrm{AME}(5,2)\rangle |0\rangle$ & $(1, 0, 10)$ \\ \cline{2-3}
                               & $|\mathrm{GHZ}(3)\rangle^{\otimes 2}$ & $(0,6,8)$\\ \cline{2-3}
                               & $|\mathrm{GHZ}(5)\rangle|0\rangle$ & $(1,10,10)$\\ \cline{2-3}
                               & $|\mathrm{GHZ}(3)\rangle |0\rangle^{\otimes 3}$ & $(3,6,14)$ \\ \hline
    \end{tabular}
    }
    \caption{Sector lengths $(S_1)$, $(S_1, S_2)$, and $(S_1, S_2, S_3)$ for states associated with certain vertices of the polytope $\RR$ defined in Theorem~\ref{Theo.RR.region}. The last three states for $N=6$ correspond to points along the edges and were found to be extremal for some other sector length (see Table~\ref{tab.3}). The states for $N=4,5,$ and $6$ are indicated in Figs.~\ref{fig1} and \ref{fig3}. }
    \label{tab.1}
\end{table}  

\section{Results for arbitrary number of qubits}
\label{Sec5.infinite}
In Sec.~\ref{Sec4.Num.Range} and Table~\ref{tab.3}, we analyzed systems with a reduced number of qubits up to $N=8$. However, in certain cases, it is possible to derive explicit bounds for certain sector lengths that hold for arbitrary values of $N$. In the following, we present several examples of such bounds. Before proceeding with the proofs, we first calculate the sector lengths for both product states and GHZ states for an arbitrary number of qubits. 

\subsection{Product and GHZ states}
We start with product states. As sector lengths are SU$(2)$ invariants, we can consider the state $\ket{0}^{\otimes N}$, for which we have
\begin{equation}
\langle \si_{\boldsymbol{\mu}}  \rangle = \prod_{\alpha=1}^N \left( \delta_{0, \mu_{\alpha}} + \delta_{3, \mu_{\alpha}}  \right) .
\end{equation}
Then, $S_m(|\psi\rangle)\equiv S_m(\rho_\psi)$ is equal to the number of Pauli strings consisting of $N-m$ 0 and $m$ 3's, \emph{i.e.}, 
\begin{equation}
\label{eq:SNproduct}
    S_m\big(\ket{0}^{\otimes N}\big) = \binom{N}{m}.
\end{equation}

We now consider the $\ket{\mathrm{GHZ}(N)}$ state. The squared expectation value $|\langle \si_{\boldsymbol{\mu}} \rangle |^2$ is equal to $1$ if $\boldsymbol{\mu}$ consists either of a vector of zeros with an even number of $3$'s, or a vector of $1$ with an even number of $2$.  In all other cases, this quantity vanishes. As a consequence, we find that $S_m = 0$ for all odd $m < N$. For even $m < N$, we have $S_m = \binom{N}{m}$, with nonzero contributions arising from Pauli strings of the form $\boldsymbol{\mu} = \boldsymbol{3}_m \boldsymbol{0}_{N-m}$, up to permutations.

For the case $m = N$, the squared expectation value $|\langle \si_{\boldsymbol{\mu}} \rangle |^2$ equals $1$ when $\boldsymbol{\mu} = \boldsymbol{1}_{N-j}\boldsymbol{2}_j$, with $j$ ranging from 0 to $N$ if $N$ is even, or from 0 to $N-1$ if $N$ is odd. These terms contribute a total of $2^{N-1}$. In addition, for even $N$, there is one more nonzero contribution from the Pauli string $\boldsymbol{\mu} = \boldsymbol{3}_N$. Therefore, we find
\begin{equation}
\label{eq:SNGHZ}
    S_N(\ket{\mathrm{GHZ}(N)}) = \left\{
    \begin{array}{lc}
        2^{N-1} +1 & \text{ for $N$ even}  \\
        2^{N-1} & \text{ for $N$ odd}
    \end{array}
    \right. .
\end{equation}
\subsection{One and two-body correlations}
\label{Subsec.1.2.body}
We found tight upper bounds for the sector lengths $S_1$ and $S_2$ that quantify the amount of one and two-body correlations. The same bounds were also derived and formally proved in Ref.~\cite{wyderka2020characterizing}. However, we observe here that these bounds follow straightforwardly from Eqs.~\eqref{Eq.Ine.R3.Eq1}--\eqref{Eq.Ine.R3.Eq3}.

First, consider Eqs.~\eqref{Eq.Ine.R3.Eq1} and~\eqref{Eq.Ine.R3.Eq2}  with $k=2$:
\begin{equation}
    \begin{aligned}
        \mbinom{N}{2} S_0 + \mbinom{N-1}{1} S_1 + \mbinom{N-2}{0} S_2 & \leq 4 \mbinom{N}{2},
    \\[2pt]
     -\mbinom{N}{2} S_0 + \mbinom{N-1}{1} S_1 - \mbinom{N-2}{0} S_2 & \leq  0 .
    \end{aligned}
\end{equation}
Subtracting the two equations, we obtain 
\begin{equation}
   S_1 \leq N.
   \label{S1N}
\end{equation}

Now, from Eq.~\eqref{Eq.Ine.R3.Eq3} with $k=3$, we obtain 
\begin{equation} 
\label{s2k3}
\mbinom{N}{3}S_0 + \mbinom{N-2}{1} S_2 \leqslant 4\mbinom{N}{3}. 
\end{equation}
Since $S_0=1$ for any $N$, it follows that
\begin{equation}
    S_2 \leq \mbinom{N}{2}.
 \label{S2N2}
 \end{equation}
One may conjecture from Eqs.~\eqref{S1N} and \eqref{S2N2} that the $S_m$ are upper bounded by binomial coefficients $\tbinom{N}{m}$. This is no longer the case for $S_3$ since for $N=4$ we have $S_3(\ket{\textrm{tetra}})=8$ [see Eq.~\eqref{tetrapsi}]. The next subsection presents more general results on these upper bounds.

The lower bound of $S_2$ for $N\leq 9$ has been obtained in the previous section and in earlier works (see Table~\ref{tab.3}). In what follows, we describe a procedure to construct states whose sector lengths $S_1$ and $S_2$ vanish simultaneously for larger values of $N$. The construction relies on the following result:
\begin{prop}
For a product state $\rho=\rho_1 \otimes \rho_2 \otimes \dots \otimes \rho_r $, it holds that
\begin{equation}
    S_m (\rho) = \sum_{\substack{j_1, \dots , j_r=0 \\ j_1 + \dots +j_r=m}}^m \prod_{i=1}^r S_{j_i} (\rho_i)  .
\end{equation}
\end{prop}
The proof follows from recursively applying
\begin{equation}
    S_m (\rho_1 \otimes \rho_2) = \sum_{j=0}^m S_j (\rho_1) S_{m-j} (\rho_2) ,
\end{equation}
which was established in Ref.~\cite{wyderka2020characterizing}. In particular, the sector length $S_1$ of a product state is given by a sum of terms, each consisting of products of $S_0$'s (equal to $1$ by normalization) and exactly one $S_1$ from its constituent states. Similarly, the contributions to $S_2$ fall into two categories: (i) products of $S_0$'s with a single $S_2$, and (ii) products of $S_0$'s with a pair of $S_1$'s. Consequently, if all constituent states $\rho_i$ of a product state $\rho$ satisfy $S_1(\rho_i)=S_2(\rho_i)=0$, then the product state itself obeys $S_1(\rho)=S_2(\rho)=0$. We now return to the problem of constructing an $N$-qubit pure state such that $S_1=S_2=0$. For $N=5,\dots,9$, explicit examples, which we denote here by $\ket{\varphi(N)}$ for simplicity, are provided in Appendix~\ref{App.states.table}. For any integer $N\geq 10$, we can write $N=5(q-1)+(5+r)$ for some natural numbers $q,r$ with $r \leq 4$. Hence, the $N$-qubit state
\begin{equation}
    \ket{\psi}= \ket{\mathrm{AME}(5,2)}^{\otimes q-1} \ket{\varphi(5+r)}
\end{equation}
is such that $S_1=S_2=0$. Therefore, for every $N \geq 5$, there exists a pure state whose sector lengths satisfy $S_1 = S_2 = 0$.

\subsection{Upper bound for \texorpdfstring{$S_{m}$}{Lg} with \texorpdfstring{$m>2$}{Lg}}
\label{Subsec.upper.Si}

The results presented in Table~\ref{tab.3} show that the inequality $S_{m} \leq \tbinom{N}{m}$ holds for all multiqubit systems with $N$ larger than some $m$-dependent $N_0$. This property was previously established in Ref.~\cite{wyderka2020characterizing}, which states the following lemma:
\begin{lemma}
\label{Lemma.Wyderka.Guhne} \cite{wyderka2020characterizing}.
    If $S_{m}(\rho) \leqslant \tbinom{N_0}{m}$ holds for all states $\rho$ of $N_0$ qubits, then for any $N\geq N_0$, $S_{m}(\rho') \leqslant \tbinom{N}{m}$ holds for all $N$-qubit states $\rho'$.
\end{lemma}
In Lemma \ref{Lemma.Wyderka.Guhne} states $\rho$ and $\rho'$ are not necessarily pure.

In this paper we find analytically the minimal values $N_0=2,3,5$ for $m =1,2,3$, respectively, and numerically $N_0=8$ for $m=4$.
The central question is whether, for each $m$, there exists a finite value of $N_0$ and if so, what the smallest such $N_0$ is, satisfying $S_{m}(\rho) \leq \binom{N_0}{m}$ for all $N_0$-qubit states $\rho$. This existence problem was first conjectured in Ref.~\cite{wyderka2020characterizing} and was recently proven in Ref.~\cite{Anschuetz2025} (see Proposition II.1 in the Supplemental Material in \cite{Anschuetz2025}), which established the bound $N_0\leq (3^k-1)/2$. This leads to the following Theorem:
\begin{theorem}
    \label{Conj.Wyderka.GuhneTHM}
\cite{Anschuetz2025}. For every $m$, there exists an $N_0\leq (3^k-1)/2$ such that $S_{m} \leqslant \binom{N_0}{m}$ for all $N_0$-qubit states. By Lemma~\ref{Lemma.Wyderka.Guhne}, this inequality then extends to all $N \geq N_0$.
\end{theorem}

The exact minimal value of $N_0$, however, is still unknown. Our results in Sec.~\ref{Sec4.Num.Range} allow us to determine $N_0$ explicitly for $m \leq 5$. For $m=1,2,$ and $3$, we recover the same minimal values of $N_0$ as reported in Ref.~\cite{wyderka2020characterizing}. By analytically maximizing $S_{m}$ under the constraints defining $\mathcal{R}$ [see Eq.~\eqref{defR}], we obtain $N_0 = 9$ for $m=4$ and $N_0 = 12$ for $m=5$. However, the numerical results in Table~\ref{tab.3} suggest that the minimal value for $m=4$ may instead be $N_0 = 8$.
\subsection{Upper bound for \texorpdfstring{$S_N$}{Lg}}
\label{Subsec.Upper.SN}
The lower bound of $S_N$ for an $N$-qubit system is $1$, and this bound is saturated by product states, as shown in Ref.~\cite{PhysRevA.94.042302}. The same reference proves that $S_N \leq 2^{N-1}$ for odd $N$, and conjectures that $S_N \leq 2^{N-1} + 1$ for even $N$. In both cases, the $\ket{\mathrm{GHZ}(N)}$ states saturate the inequality. This conjecture has been rigorously proven in Ref.~\cite{Eltschka2020}, leading to the following theorem:
\begin{theorem}
\label{Theo.2}
\cite{PhysRevA.94.042302,Eltschka2020}
For an $N$-qubit pure state, the sector length $S_N$ has a minimum value of 1. Its maximum value is
$$
\max S_N = 
\begin{cases}
2^{N-1}, & \text{if } N \text{ is odd}, \\
2^{N-1} + 1, & \text{if } N \text{ is even}.
\end{cases}
$$
The minimum and maximum are attained by product states and GHZ states, respectively.
\end{theorem}

Here, we give an alternative proof of this result. Our strategy is the following: find an upper bound for $S_N$ within some polytope containing $\mathcal{S}$, and then show that this upper bound is reached within $\mathcal{S}$, i.e.~saturated by some quantum state.

Finding an upper bound for $S_N$ under the constraint that the vector $\mathbf{S}$ belongs to a polytope is a typical linear programming problem. In our case, we observed numerically that the maximum of $S_N$ is reached when optimizing only under the constraints \eqref{symetrieS} for $1\leq k\leq N/2$ and \eqref{Eq.Ine.R3.Eq3} for odd $k$. The linear programming problem in the variables $S_1,...,S_N$ can be expressed as follows:\\
\begin{itemize}
    \item find $\max(S_N)$ 
\item under the constraints $S_m\geq 0$, $m=1,...,N$
\item with inequality constraints for $q=1,...,N/2-1$
\begin{equation}
    \sum_{m=1}^{q}\mbinom{N-2m}{2q+1-2m}S_{2m}\leq \left(2^{2q}-1\right)\mbinom{N}{2q+1},
 \label{inequalities}
 \end{equation}
corresponding to \eqref{Eq.Ine.R3.Eq3} for $k=2q+1$,
\item and equality constraints for $k=0,...,N/2-1$
\begin{multline}
    \frac{1}{2^{N-k}}\sum_{m=1}^{N-k}\mbinom{N-m}{k}\,S_{m}- 
    \frac{1}{2^k}\sum_{m=1}^{k}\mbinom{N-m}{N-k}\,S_{m} =\\
   \left(\frac{1}{2^k}-\frac{1}{2^{N-k}}\right)\mbinom{N}{k}.
   \label{equalities}
\end{multline}
\end{itemize}
This linear problem takes the form of a primal problem over the variables $S_m$ \cite{boyd2004convex}: 
\begin{itemize}
    \item find $\max(\sum_m c_m S_m)$ 
    \item under the constraints $S_m\geq 0$, $m=1,...,N$, 
    \item with $\sum_m a_{qm}S_m\leq b_q$ for $q=1,...,N/2-1$ 
    \item and $\sum_m a'_{k m}S_m= b'_k$ for $k=0,...,N/2-1$,
\end{itemize}
with 
\begin{equation}
\label{defca}
\begin{aligned}
   & c_{m}=\delta_{m,N}\\[2pt]
   & a_{qm}=\tbinom{N-m}{2q+1-m}\delta_{m\textrm{ even}}\\[2pt]
   & a'_{k m}=\tfrac{1}{2^{N-k}}\tbinom{N-m}{k}-\tfrac{1}{2^{k}}\tbinom{N-m}{N-k}\\
   & b_{q}=\left(2^{2q}-1\right)\tbinom{N}{2q+1}\\
   & b'_k=\left(\tfrac{1}{2^k}-\tfrac{1}{2^{N-k}}\right)\tbinom{N}{k}.
\end{aligned}
\end{equation}
To this primal problem corresponds the following dual problem over the variables $y_q,y'_k$, corresponding to constraints \eqref{equalities} and \eqref{inequalities}, respectively:
\begin{itemize}
    \item find $\min(\sum_q b_q y_q+\sum_k b'_k y'_k)$ 
\item under the constraints $y_q\geq 0$, $q=1,...,N/2-1$,
\item and inequality constraints for $m=1,...,N$
\begin{equation}
\label{ineq_dual}
   \sum_{q=1}^{N/2-1} a_{qm}y_q+\sum_{k=0}^{N/2-1} a'_{k m}y'_k\geq c_m.
\end{equation}
\end{itemize}
The constraints defining the primal problem are a subset of the constraints defining $\RR\supset\mathcal{S}$. Therefore, all points in $\mathcal{S}$ are feasible points of the primal problem and, by virtue of the weak duality theorem of linear programming, provide a lower bound for the dual problem. This is the case of the state $\ket{\mathrm{GHZ}(N)}$, for which Eq.~\eqref{eq:SNGHZ} gives $S_N=2^{N-1}+1$. From the strong duality theorem, to conclude the proof, it suffices to show that there are feasible values of $y_q,y'_k$ which achieve that lower bound.

We take the following values:
\begin{equation}
\label{values_dual}
    \begin{array}{rll}
         y_q&=0&1\leq q\leq \frac{N}{4}-1  \\[2pt]
         y_{q}&=2^{1-N/2}&q=\frac{N-2}{4} \textrm{ if } N\!\!\!\!\!\mod 4=2\\[2pt]
     y_q&=2^{-2q+1}&\lfloor\frac{N+2}{4}\rfloor\leq q\leq \frac{N}{2}-1\\[2pt]
         y'_{0}&=2^{N}&\\[2pt]
         y'_k&= \frac{2^{N-k}-2^k+1}{(-1)^k }-1,&\quad 1\leq k\leq \frac{N}{2}-1.
    \end{array}
\end{equation}
The values \eqref{values_dual} obviously follow the trivial constraints $y_q\geq 0$. As for the other inequality constraints, the left-hand side of Eq.~\eqref{ineq_dual} reads as
\begin{equation}
Q_m=\sum_{q=1}^{N/2-1} a_{qm}y_q+\sum_{k=0}^{N/2-1} a'_{k m}y'_k.
\end{equation}
In Appendix \ref{Qpositif1} we show that $Q_m\geq c_m$, which shows that the values \eqref{values_dual} yield a feasible point for the dual problem. 
We show in Appendix \ref{Qpositif2} that at this point we have
\begin{equation}
\label{eq_min}
\sum_{q=1}^{N/2-1} b_q y_q+\sum_{k=0}^{N/2-1} b'_k y'_k=2^{N-1}+1,
\end{equation}
and thus $2^{N-1}+1$ is an upper bound of the primal problem. This proves that $S_N\leq 2^{N-1}+1$ for any pure state, with equality for the $\ket{\mathrm{GHZ}(N)}$ state.

\subsection{The overlap \texorpdfstring{$R_{\rho}$}{Lg}}
Since the quantity $\Tr(\rho_A^2) + R_{\rho_A}$ played a central role in deriving our inequalities for the sector lengths, we now examine its extremal values. In particular, we investigate whether the bounds given in Proposition~\ref{Prop.1} are saturated. Without loss of generality, we consider a bipartition $A|\bar{A}$, where $A$ consists of the first $k$ qubits. The case $k = 1$ is a bit special because the Hilbert space of $A$ is two dimensional. In this setting, any orthonormal basis $\{ \ket{\phi_1}, \ket{\phi_2} \}$ must satisfy $\ket{\phi_2} \propto \ket{\tilde{\phi}_1}$ since the orthogonal complement of $\ket{\phi_1}$ is one dimensional, and for an odd number of qubits, $\langle \phi_1 | \tilde{\phi}_1 \rangle = 0$ (see Subsection~\ref{purestateIneqs}). As a result, using Eq.~\eqref{rrhoatrrhoa}, we find that for $|A| = 1$,
\begin{equation}
\label{Eq.Sum.rho.R.k1}
\Tr(\rho_{A}^2) + R_{\rho_{A}} =1 ,
\end{equation}
which we also found numerically for small values of $N$ (see Table~\ref{tab.4}).

For $k = |A| > 1$, the upper bound of $\Tr(\rho_A^2) + R_{\rho_A}$ depends on the parity of $k$. When $k$ is odd, the bound is attained by product states with $\Tr(\rho_A^2) = 1$ and $R_{\rho_A} = 0$. For even $k$, the maximum is achieved by a state of the form $\ket{\psi_A} \otimes \ket{0}^{\otimes N_B}$, where $\ket{\psi_A}$ satisfies $\ket{\tilde{\psi}_A} \propto \ket{\psi_A}$. An example is the Dicke state $\ket{D_k^{(k/2)}}$ defined in Eq.~\eqref{Eq.Dicke}, for which $\Tr(\rho_A^2) = 1$ and $R_{\rho_A} = 1$. This maximum value can be interpreted as a compromise between the purity of the reduced state $\rho_A$ and the overlap $R_{\rho_A}$, when the sum is optimized.

For the minimum of $\Tr(\rho_A^2) + R_{\rho_A}$, Table~\ref{tab.4} lists examples of optimal states obtained numerically for systems with up to $N = 10$ qubits and various values of $k$. The lower bound given in Proposition~\ref{Prop.1} is saturated when $N_A > N_{\bar{A}}$, \emph{i.e.}, when $2k > N$. For $k \leq N - k$ and $k \neq 1$, numerical results consistently yield a minimum of $2^{-k + 1}$. It should be noted that the case $N = 10$ with $k = 5$ provides the first example in which the contributions of $\Tr(\rho_A^2)$ and $R_{\rho_A}$ are neither completely balanced nor entirely concentrated in a single term. 
All these observations can be summarized in the following conjecture:
\begin{conjecture}
\label{Conj.3}
Let $A|\bar{A}$ be a bipartition of an $N$-qubit system into $k=|A|$ and $N-k=|\bar{A}|$ qubits. Then, the minimum of $\mathrm{Tr}(\rho_A^2) + R_{\rho_A}$ taken over all pure states $|\psi\rangle$ of the $N$-qubit system, is given by
\begin{equation}
    \min_{|\psi\rangle} \big(\mathrm{Tr}(\rho_A^2) + R_{\rho_A}\big) = 
    \begin{cases}
\: 2^{k-N}, & \text{if } 2k > N, \\[2pt]
\: 2^{-k + 1}, & \text{if } 2k \leq N.
\end{cases}
\end{equation}
\end{conjecture}

\section{Applications}
\label{Sec6.App}
Sector lengths can be used to define entanglement measures for pure states (see Appendix~\ref{App.multipartite.concurrence} for more details). In this section, we explore several applications of sector lengths related to these entanglement measures and to quantum coding theory.
\subsection{Linear entropy of entanglement and shadow enumerators}
Sector lengths are directly related to the linear entropy of entanglement of a pure state across a bipartition $A|\bar{A}$, defined as
\begin{equation}
E^A_L(\rho) \equiv 2 \left[1-
\Tr \left( \rho_A^2 \right)
\right] .
\end{equation}
Using Eqs.~\eqref{trArA} and \eqref{sumaSi}, the average linear entropy of entanglement over all bipartitions of $k$ and $N-k$ qubits, $\overline{E}_L^{k} (\rho) \equiv \sum_{|A|=k} E^A_L(\rho)
$, is indeed equal to
\begin{equation}
\begin{aligned}    
\overline{E}_L^{k} (\rho) 
    = {}&
    2 \mbinom{N}{k} \left[ 
    1-  \frac{1}{2^k} \sum_{m=0}^k \frac{\mbinom{k}{m} }{\mbinom{N}{m}} S_m(\rho) 
    \right] .
    \end{aligned}
\end{equation}
The quantity $\overline{E}_L^k(\rho)$ vanishes if and only if the state is a product state. This is because $\overline{E}_L^k(\rho)$ is a positive linear combination of entanglement measures. Hence, $\overline{E}_L^k = 0$ implies that the state is separable across all bipartitions $A|\bar{A}$ with $|A| = k$, which is equivalent to being a product state. On the other hand, for a $k$-uniform state, that is, a state that is maximally entangled across every bipartition with up to $k$ qubits in subsystem $A$, it holds that
\begin{equation}
x_{\bmu_{m} \mathbf{0}_{N-m} }= \delta_{\bmu_{m} \mathbf{0}_{m}}
\end{equation}
for any $m$-index vector $\bmu_{m}$. This is equivalent to
\begin{equation}
    S_{m} =0  \quad  \forall \, m \leq k .
\end{equation}
Such $k$-uniform states can exist only for $k \leq \lfloor N/2 \rfloor$. When $k = \lfloor N/2 \rfloor$, they are known as absolutely maximally entangled (AME) states and denoted by $\ket{\mathrm{AME}(N,2)}$~\cite{GISIN19981,PhysRevA.86.052335}, where the number 2 is related with the dimension of each subsystem. Thus, the maximum possible average linear entropy of entanglement is attained only by $k$-uniform states.

Other relevant quantities for quantifying entanglement are the shadow enumerators $\Shpsi_g$, which are linear combinations of sector lengths [see Eq.~\eqref{Eq.Kravchuk}]. These quantities have recently been used in the experimental characterization of entanglement~\cite{miller2024}. As shown in Appendix~\ref{App.Doublecopy}, each shadow enumerator corresponds to the expectation value of a projector operator acting in the double-copy Hilbert space. Furthermore, as we discuss in Appendix~\ref{App.multipartite.concurrence}, some of them can also be interpreted as 
multipartite concurrences. For pure states, the shadow enumerators satisfy $S_g^{(e)}(\rho) =0$ for $g-N$ odd and $\sum_{g=0}^{N} S_g^{(e)}=2^N$ (see Appendix~\ref{App.Doublecopy}).

For $N \leq 5$, we showed in Sec.~\ref{Sec5.infinite} that the numerical range $\Scal$ forms a polytope. It follows that the extremal values of any physical quantity expressed as a linear combination of $S_m$ are reached by quantum states necessarily associated with the vertices of this polytope. More generally, the maximum or minimum of any monotonic function of such a linear combination will be reached at one of the vertices of the polytope\footnote{If the optimum is reached at several vertices, the complete extremal set corresponds to the convex hull of these vertices in the space $S_m$.}. As a concrete example, for $N = 4$ and $5$, the quantities $\overline{E}_L^{g}$ and $S_g^{(e)}$ with $g < N$ can be extremized by evaluating them over the finite set of extremal states from Table~\ref{tab.1}, with the corresponding maximizers and minimizers summarized in Table~\ref{tab.2} (including results for $N = 6$, when available).

As discussed above, product states minimize $\overline{E}_L^k$, while for $N = 5$ and $6$, this quantity is maximized by AME states. For $N = 4$, however, an $|\mathrm{AME}(4,2)\rangle$ state does not exist. In this case, the state $\ket{\text{tetra}}$ maximizes $\overline{E}_L^2$; further details about this state can be found in Appendix~\ref{App.states.table}.

For $N = 4$, the shadow enumerators~\eqref{Eq.Kravchuk} give
\begin{equation}
    \Shpsi_0 = \frac{1}{4} \left(-2-S_1+ S_2 \right) , \quad 
    \Shpsi_2 = \frac{1}{2} \left(18- 3S_1-S_2 \right) ,
\end{equation}
and for $N = 5$
\begin{equation}
    \Shpsi_1 = \frac{1}{2} \left(-2 S_1+S_2\right) , \quad
    \Shpsi_3 = -2 S_1-S_2+20 .
\end{equation}
Note that $\Shpsi_3$ is maximized not only by $|\mathrm{AME}(5,2)\rangle$ states, but also by the five-qubit GHZ state, and that shadow enumerators are not always minimized by product states.

For $N \geq 6$, our results are not sufficient \emph{a priori} to completely characterize the structure of $\Scal$; however, we can search for extrema over the larger set $\RR$. Since $\Scal \subset \RR$, if an extremum is reached at a vertex of $\RR$ that corresponds to a realizable quantum state, then that state is guaranteed to be optimal. We apply this approach for $N = 6$ to both the average linear entanglement entropy and the shadow enumerators. The linearly independent shadow enumerators in this case are
\begin{equation}
\begin{aligned}    
    \Shpsi_0 ={}&\frac{1}{8} \left(8+2 S_1-S_3 \right) , 
    \\ \Shpsi_2 ={}& \frac{1}{8} \left( 
    -30 S_1+8 S_2+3 S_3
    \right) ,
    \\ \Shpsi_4 ={}&
\frac{1}{8} \left(
360-10 S_1-16 S_2-3 S_3
    \right) .    
\end{aligned}
\end{equation}
We find that product states minimize all three shadow enumerators, though some of these minima are also attained by other states. The state $\ket{\mathrm{AME}(6,2)}$ is found to maximize both $\Shpsi_0$ and $\Shpsi_4$. Additionally, using our inequalities, we can analytically confirm that $\ket{\mathrm{GHZ}(6)}$ maximizes $\Shpsi_0$. Numerical results further suggest that it also maximizes $\Shpsi_2$, although this cannot be concluded from our inequalities alone.
\begin{table}[t!]
    \centering
    \renewcommand{\arraystretch}{1.5}
    \begin{tabular}{|c|c|c|c|}
        \hline
        $N$ & State  & $\,$ Minimum $\,$ & $\,$ Maximum $\,$
        \\ 
        \hline\hline
        \multirow{3}{*}{$\;4\;$} & $|\mathrm{tetra}\rangle$ & $\Shpsi_0$  & $\overline{E}_L^{1} , \overline{E}_L^{2} , \Shpsi_2 $ \\ \cline{2-4}
                               & $|\mathrm{GHZ}(4)\rangle$   &  & $\overline{E}_L^{1} , \Shpsi_0  $ \\ \cline{2-4}
                               & $|0\rangle^{\otimes 4}$ & $\overline{E}_L^{1} , \overline{E}_L^{2} , \Shpsi_0 , \Shpsi_2 $ & \\ \hline\hline
        \multirow{3}{*}{$\;5\;$} & $|\mathrm{AME}(5,2)\rangle$ & $ \Shpsi_1 $ & $\overline{E}_L^{1} , \overline{E}_L^{2} , \Shpsi_3$ 
        \\ \cline{2-4}
                               & $|\mathrm{GHZ}(5)\rangle$   &  & $\overline{E}_L^{1} , \Shpsi_1$ \\ \cline{2-4}
                               & $|0\rangle^{\otimes 5}$ & $\overline{E}_L^{1} , \overline{E}_L^{2} , \Shpsi_1 ,  \Shpsi_3 $ & \\ \hline\hline
        \multirow{10}{*}{$\;6\;$} & \multirow{2}{*}{$|\mathrm{AME}(6,2)\rangle$} & \multirow{2}{*}{$\Shpsi_2$} & $\overline{E}_L^{1} , \overline{E}_L^{2} , \overline{E}_L^{3}    $\\
                               &  &  & $\Shpsi_0 , \Shpsi_4   $\\\cline{2-4}
                               & $|\psi_6\rangle^{\eqref{Eq.psi6}}$       & $ \Shpsi_0$ & $\overline{E}_L^{1} , \overline{E}_L^{2}$\\ \cline{2-4}
                               & $|\mathrm{GHZ}(6)\rangle$  &  & $\overline{E}_L^{1}, \Shpsi_0$, $\mathit{\Shpsi_2}\footnote{The optimality of $\ket{\mathrm{GHZ}(6)}$ for $\Shpsi_2$ relies only on a global numerical optimization over the pure states space, not on our inequalities.}$\\ \cline{2-4}
                               & \multirow{2}{*}{$|0\rangle^{\otimes 6}$} & $\overline{E}_L^{1} , \overline{E}_L^{2} , \overline{E}_L^{3}  $ & \\ 
                               &  & $ \Shpsi_0 ,  \Shpsi_2 ,  \Shpsi_4 $ & \\ \cline{2-4}
                               & $\;|\mathrm{AME}(5,2)\rangle |0\rangle\;$ &  $\Shpsi_0 ,  \Shpsi_2$ & \\
                               \cline{2-4}
                               & $|\mathrm{GHZ}(3)\rangle^{\otimes 2}$ & $ \Shpsi_0 $&$\overline{E}_L^{1} $\\ \cline{2-4}
                               & $|\mathrm{GHZ}(5)\rangle|0\rangle$ & $\Shpsi_0$ &\\ \cline{2-4}
                               & $|\mathrm{GHZ}(3)\rangle |0\rangle^{\otimes 3}$ & $\Shpsi_0 , \Shpsi_2$ & \\ \cline{2-4}
                               \hline
    \end{tabular}
    \caption{States lying on the boundary of the polytope $\mathcal{R}$ and quantities they extremize among the average linear entanglement entropy and the shadow enumerators, $\overline{E}_L^{g} $ and $\Shpsi_g$, respectively.}
    \label{tab.2}
\end{table}
\subsection{Quantum coding theory}
Let us now discuss the connection between our results and quantum-error correction. For completeness, we briefly summarize key concepts of \emph{quantum error-correcting codes} (QECCs) in the context of sector lengths restricted to qubits, following the references~\cite{681316,PhysRevA.69.052330}. Any error operator can be expressed as a linear combination of \emph{local error operators}, which are simply Pauli strings $\si_{\bm{\mu}}$. Consequently, we focus on errors of the form $E=\si_{\bm{\mu}}$.

A \emph{quantum code} $\mathcal{Q}$ of dimension $K$ is a vector subspace of the Hilbert space of $N$ qubits, $\left(\mathbb{C}^2\right)^{\otimes N}$. An error operator $E$ is said to be \emph{detectable by} $\mathcal{Q}$ if
$$
\bra{\psi} E \ket{\psi} = c(E) \qquad \textrm{ for all} \ket{\psi} \in \mathcal{Q},
$$
where $c(E)$ is a constant depending only on $E$.

A quantum code $\mathcal{Q}$ for qubits, specified by the triple $((N,K,d))$, is a subspace of dimension $K$ that detects all local errors with $\wt(\sigma_{\bm{\mu}}) <d$. Such a code is referred to as a $((N,K,d))$ QECC. The code is called \emph{pure} if, for any local error $E$, the expectation value satisfies $c(E) =2^{-N} \Tr (E)$. In our context, we focus on a single pure state, corresponding to a one-dimensional subspace ($K=1$). QECCs of the form $((N, 1, d))$ are known as \emph{self-dual} codes and are, by convention, always considered pure. This basic theory allows us to highlight a link between QECCs and sector lengths.
\begin{prop}
\cite{681316,PhysRevA.69.052330}:  A pure state $\ket{\psi}$ of $N$ qubits has $S_m=0$ if and only if it defines a (pure) $((N,1,m+1))$ QECC. 
\end{prop} 
In quantum coding theory, a more quantitative way to determine whether a code detects local errors of weight $m$ is through the Shor-Laflamme enumerators, which are defined in terms of the projector $M_{\mathcal{Q}}$ on the code subspace $\mathcal{Q}$ by~\cite{PhysRevLett.78.1600,Rains:796376}
\begin{equation}
\label{Eq.Shor.Laflamme.bis}
\begin{aligned}    
    A_{m} (M_{\mathcal{Q}},M_{\mathcal{Q}}) \equiv{}& \sum_{\wt(\si_{\bm{\mu}})=m} \Tr \left( \si_{\bm{\mu}} M_{\mathcal{Q}} \right)^2,
\\
    B_{m} (M_{\mathcal{Q}},M_{\mathcal{Q}}) \equiv{}& \sum_{\wt(\si_{\bm{\mu}})=m} \Tr \left( \si_{\bm{\mu}} M_{\mathcal{Q}} \si_{\bm{\mu}} M_{\mathcal{Q}} \right),
\end{aligned}
\end{equation}
for $m=1, \dots,  N$. In the special case where $K = 1$ (i.e., the code consists of a single state), the projector becomes $M_{\mathcal{Q}} = \rho = \ket{\psi}\bra{\psi}$. In this case, the Shor-Laflamme enumerators reduce to the sector lengths: $A_m(\rho, \rho) = B_m(\rho, \rho) = S_m(\rho)$. For completeness, we present this result along with the more general framework of Shor-Laflamme enumerators~\cite{PhysRevLett.78.1600} and their connection to sector lengths in Appendix~\ref{App.Shor.Laflamme}. We also discuss their relation to the MacWilliams identity and shadow inequalities.
\section{Conclusions}
\label{Sec7.Conclusions}
In this work, we studied the set $\mathcal{S}$ of admissible sector lengths $\mathbf{S} = (S_1, \dots , S_N)$ for pure $N$-qubit states. By combining known constraints on multipartite quantum correlations with the monogamy inequalities derived in this work, we constructed a polytope $\mathcal{R}$ containing $\mathcal{S}$ (see Theorem~\ref{Theo.RR.region}). For $N \leq 5$ qubits, we showed that this polytope precisely defines the boundary of $\mathcal{S}$.
The states corresponding to the edges of $\mathcal{R}$ were identified, and those corresponding to the vertices are listed in Table~\ref{tab.1}. Extending our analysis to six-qubit systems revealed a significant increase in complexity, with numerical evidence suggesting that $\mathcal{R} \neq \mathcal{S}$ in this case. The origin of additional constraints on sector lengths defining $\mathcal{S}$ for $N \geq 6$ remains to be determined, as does the more general question of whether the set $\mathcal{S}$ continues to form a polytope for larger $N$. 

We also systematically identified, numerically or analytically, pure states that extremize sector lengths $S_m$ (\emph{i.e.}, lie on the boundary of $\mathcal{S}$) for all $m$ for $N\leq 8$. These states, listed in Table~\ref{tab.3}, include known exceptional states as well as previously unknown states. Those highlighted in gray have been analytically verified as optimal by saturating inequalities defining $\RR \supset \mathcal{S}$. Our inequalities also provided an alternative proof for a conjecture on the maximum value of $S_N$ (see Theorem~\ref{Theo.2}) from Ref.~\cite{PhysRevA.94.042302}, and allowed us to find analytically the minimum number of qubits $N_0$ for which Theorem~\ref{Conj.Wyderka.GuhneTHM} holds, namely, $N_0=9$ for $S_4$ and $N_0=12$ for $S_5$.

The search for new monogamy inequalities is essential to refine the geometric understanding of the set $\mathcal{S}$. To this end, we examined the sum of the purity and the overlap of the reduced states, $\Tr(\rho_A^2) + R_{\rho_A}$, and found that some extremal states do not saturate our bounds (see Conjecture~\ref{Conj.3}), suggesting that there is still potential to find further and sharper inequalities. A complete characterization of the set $\mathcal{S}$ is valuable for identifying states that optimize physical quantities expressible only in terms of sector lengths. When such a quantity is a monotonic function of a linear combination of $S_m$, the search space reduces to the boundary of $\Scal$, or if $\mathcal{S}$ is a polytope, its vertices. Leveraging this, we determined states extremizing the average linear entropy of entanglement and the quantum shadow enumerators for $N = 4, 5$ and $6$ (see Table~\ref{tab.2}). Both quantities, which are examples of generalized concurrences, depend linearly on sector lengths. Additionally, we showed that the shadow enumerators can be interpreted as the expectation value of a projector in the double-copy Hilbert space.

In summary, our results once again highlight the complex structure of multipartite quantum correlations, with a notable increase in complexity at $N = 6$, suggesting the existence of stricter constraints that remain to be found. The extremal states identified here could constitute key resources in the fields of quantum entanglement and quantum information, motivating further studies on their properties and potential generalizations.
\section*{Acknowledgements}
E.S.E.\ acknowledges support from the postdoctoral fellowship of the IPD-STEMA program of the University of Liège (Belgium). J.M.\ and E.S.E.\ acknowledge the FWO and the F.R.S.-FNRS for their funding as part of the Excellence of Science programme (EOS Project No.\ 40007526). O.G.\ thanks T.\ Paterek for valuable discussions. We also thank N.\ Wyderka for helpful correspondence.
\begin{appendix}
\section{Derivation of Eqs.~\texorpdfstring{\eqref{symetrieS}}{Lg} and \texorpdfstring{\eqref{sumaSi}}{Lg}}
\label{App.first.Eqs}
Let us first prove Eq.~\eqref{sumaSi}. We start by considering the simplest case, when subsystem $A$ contains $k=3$ qubits.
\subsection{Case \texorpdfstring{$k=3$}{Lg}}
Consider a $3$-qubit mixed state $\rho_A$. If $\rho_A$ is the reduced state of a larger pure quantum state $\rho_\psi=|\psi\rangle\langle\psi|$ of $N$ qubits, its coordinates correspond to the variables $x_{\bmu}$, where the multi-indices $\bmu$ have at least $N-3$ zeros (see Eqs.~\eqref{tensorrepmixed}--\eqref{Eq.coord.reduced.mixed}). For example, if subsystem $A$ corresponds to the three first qubits, its variables are the $x_{\mu_1 \mu_2 \mu_3 \bfo_{N-3}}$ of the original pure state of the $N$-qubit system. We now calculate the sector length $S_1$ of $\rho_A$, which is
\begin{equation}
      S_1(\rho_A)=\sum_{a=1}^3 \left( x_{a00\bfo_{N-3}}^2+ x_{0a0\bfo_{N-3}}^2 + x_{00a\bfo_{N-3}}^2 \right) .
\end{equation}
Summing over all $\tbinom{N}{k}$ bipartitions $A\vert \bar{A}$, where $\bar{A}$ is the complementary subset of $A$, each $x^2_{a00\bfo_{N-3}}$ appears in all triplets involving the first qubit, that is, $\tbinom{N-1}{2}$ times. Therefore
\begin{equation}
\label{S1k3}
\begin{aligned}    
    &\sum_{|A| =k} S_1(\rho_A)\\
    &=
    \mbinom{N-1}{2}\sum_{a=1}^3 \left(x^2_{a00...0}+x^2_{0a0...0}+\ldots+x^2_{0...0a}\right)
    \\
    &= \mbinom{N-1}{2}\,S_1(\rho_\psi) .
\end{aligned}
\end{equation}
Similarly,
\begin{equation}
      S_2(\rho_A)=\sum_{a,b=1}^3 \left(x^2_{ab0\bfo_{N-3}}+x^2_{a0b\bfo_{N-3}}+x^2_{0ab\bfo_{N-3}}\right)
\end{equation}  
and each $x^2_{ab0\bfo_{N-3}}$ appears $\tbinom{N-2}{1}=N-2$ times in the sum over $A$, thus
\begin{equation}
\label{S2k3}
\begin{aligned}    
   &  \sum_{|A|=k} S_2(\rho_A) \\
    & 
    =(N-2)\sum_{a,b=1}^3 \left(x^2_{ab0...0}+x^2_{a0b0...0}+\ldots+x^2_{0...0ab}\right)
    \\
    &
    =\mbinom{N-2}{1}\,S_2(\rho_\psi).
\end{aligned}
\end{equation}
\subsection{General \texorpdfstring{$k$}{Lg}}
Equations~\eqref{S1k3} and \eqref{S2k3} are easily generalized to arbitrary values of $k$. 
If $A$ corresponds to the $k$ first qubits, $S_{m}(\rho_A)$ with $m \leq k$ is given by
\begin{multline}
\label{sirhoa}
S_{m}(\rho_A)= \hspace*{-0.3cm} \sum_{a_1,...,a_{m}=1}^3 \Big(
x^2_{a_1a_2...a_{m}0..0\bfo_{N-k}}+\ldots
\\
+x^2_{00a_1a_2...a_{m}\bfo_{N-k}} \Big) .
\end{multline}
We then sum over all $\tbinom{N}{k}$ bipartitions $A\vert \bar{A}$, where $A$ consists of $k$ qubits. Each term $x^2_{a_1a_2...a_{m}\bfo_{k-m}\bfo_{N-k}}$ appears in every $k$-qubit subsystem that includes the first $m$ qubits, which occurs $\tbinom{N-m}{k-m}$ times. However, it appears only once in $S_{m}(\rho_\psi)$. This is true for all terms in Eq.~\eqref{sirhoa}, which leads to Eq.~\eqref{sumaSi} for $m \leqslant k$. 

Equation~\eqref{symetrieS} is then obtained by expressing the purity of the reduced state $\rho_A$ as
\begin{equation}
\label{Eq.A6}
    \Tr (\rho_A^2) = \frac{1}{2^k} \sum_{m=1}^k S_{m} (\rho_A) .
\end{equation}
Combining this with Eq.~\eqref{sumaSi}, we obtain
\begin{equation}
\label{Eq.A7}
 \sum_{|A|=k} \Tr \left( \rho_A^2 \right) = \frac{1}{2^k} \sum_{m=1}^k \mbinom{N-m}{k-m} S_{m}(\rho_\psi) .
\end{equation}
Since $\Tr (\rho_A^2) = \Tr (\rho_{\bar{A}}^2)$ holds for any pure state, we have that
\begin{equation}
    \sum_{|A|=k} \Tr \left( \rho_{A}^2 \right) 
    = 
    \sum_{|\bar{A}|=k} \Tr \left( \rho_{\bar{A}}^2 \right) 
    ,
\end{equation}
which directly leads to Eq.~\eqref{symetrieS}.
\section{Shor-Laflamme enumerators and sector lengths}
\label{App.Shor.Laflamme}
We describe the connection between sector lengths and some concepts that appear in quantum coding theory~\cite{PhysRevLett.78.1600,Rains:796376}. For any two Hermitian operators $M_1$ and $M_2$, the \emph{Shor-Laflamme enumerators} (or weights) are defined as\footnote{We adopt the definition of the Shor-Laflamme enumerators from Ref.~\cite{Rains:796376}, which agrees with the original formulation up to a global factor~\cite{PhysRevLett.78.1600}.}
\begin{equation}
\label{Eq.Shor.Laflamme}
\begin{aligned}    
    A_{m} (M_1, M_2) \equiv{}& \sum_{\wt(\si_{\bm{\mu}})=m} \Tr (\si_{\bm{\mu}} M_1) \Tr (\si_{\bm{\mu}} M_2)  ,
\\
    B_{m} (M_1, M_2) \equiv{}& \sum_{\wt(\si_{\bm{\mu}})=m} \Tr (\si_{\bm{\mu}} M_1 \si_{\bm{\mu}} M_2)  .
\end{aligned}
\end{equation}
We define two-variable polynomials using these enumerators 
\begin{equation}
\begin{aligned}
    A(x,y) = & \sum_{0\leqslant i \leqslant N} A_{m} (M_1,M_2) x^{N-m} y^{m} , 
\end{aligned}
\end{equation}
and analogously for $B(x,y)$. Theorem 7 of Ref.~\cite{681316} (first proved in Ref.~\cite{PhysRevLett.78.1600}) establishes a duality relation between $A(x,y)$ and $B(x,y)$, namely
\begin{equation}
\begin{aligned}    
    B(x,y) \equiv{}& A\left(  \frac{x+3y}{2} , \frac{x-y}{2}\right) , 
    \\ 
    A(x,y) \equiv{}& B\left(  \frac{x+3y}{2} , \frac{x-y}{2}\right) . 
\end{aligned}
\end{equation}
These identities are the quantum analog of the \emph{MacWilliams identities} used in classical coding theory over $\mathrm{GF}(4)$~\cite{macwilliams1977theory}.

In Ref.~\cite{Rains:796376}, Rains introduced another enumerator inspired by classical shadow codes~\cite{Conway:59931}. These \emph{quantum shadow enumerators}, denoted by $\Sh_{m}$, are given by
\begin{equation}
\Sh_{m} (M_1 , M_2) = B_{m} (M_1 , \tilde{M}_2), 
\end{equation}
where $\tilde{M_2} = \sigma_y^{\otimes N} M^{*}_2 \sigma_y^{\otimes N} $, and $M^{*}_2$ is the complex conjugation of $M_2$. Moreover, the associated polynomial $\Sh(x,y)$ satisfies~\cite{Rains:796376}
\begin{equation}
   \Sh(x,y) = A\left(  \frac{x+3y}{2} , \frac{y-x}{2}\right) .    
\end{equation}
Theorem 10 of Ref.~\cite{Rains:796376} shows that
\begin{equation}
\label{Eq.Shadow}
    \Sh_{m} = \frac{1}{2^N} \sum_{i=0}^N (-1)^i K_{m}(i,N) A_i\geqslant 0 .
\end{equation}
The inverse relation of Eq.~\eqref{Eq.Shadow} is given by
\begin{equation}
    \begin{aligned}
        \sum_{i=0}^N  K_m(i,N) C_{i} ={}&
        \frac{1}{2^N} \sum_{i,j=0}^N 
        (-1)^j K_m(i,N) K_{i}(j,N) A_j
        \\
        ={}&   2^N (-1)^m A_m \, ,
    \end{aligned}
\end{equation}
where the second equality follows from the orthogonality relation of the Kravchuk polynomials (Corollary 2.3 in Ref.~\cite{412678})
\begin{equation}
\label{Eq.OR.Kravchuk}
     \sum_{i=0}^N 
        K_{m'}(i,N) K_{i}(m,N)
        = 4^N \delta_{m,m'}  .
\end{equation}
We now establish the connections between the Shor-Laflamme enumerators, the sector lengths $S_{m}$, and the shadow enumerators $\Shpsi$ for quantum states. For $M_1=M_2 = \rho$, we have that $A_{m}(\rho,\rho)=S_{m}(\rho)$ and $\Sh_{m}(\rho,\rho)=\Shpsi_{m}(\rho)$, which leads to Eq.~\eqref{Eq.Kravchuk}. 
If we use instead $M_1= \rho = \tilde{M_2}$, we obtain the inequality
\begin{equation}
\label{Eq.New1}
    \frac{1}{2^N} \sum_{m=0}^N K_{m'}(m,N) \underbrace{A_{m}(\rho , \rho)}_{S_{m}(\rho)}
    \geqslant 0 ,
\end{equation}
where we have used $A_m(M_1 ,\tilde{M}_2) = (-1)^m A_m(M_1 ,M_2)$ which follows from the properties of Pauli strings. Alternatively, we can use $M_1 = M_2 = \rho_A$ with $\rho_A$ the $k$-qubit reduced state obtained after tracing over a subsystem $\bar{A}$. Thus, we obtain another family of inequalities
\begin{equation}
    \frac{1}{2^{k}} \sum_{i=0}^{k} (-1)^i K_{m}(i,k) A_i(\rho_A , \rho_A) \geqslant 0 . 
\end{equation}
Summing over all the partitions of the same size (as in Eqs.~\eqref{Eq.A6} and~\eqref{Eq.A7}) yields
\begin{equation}
\label{Eq.New2}
    \frac{1}{2^{k}} \sum_{i=0}^{k} (-1)^i K_{m}(i,k) \mbinom{N-i}{k-i} \underbrace{A_i(\rho , \rho)}_{S_i(\rho)} \geqslant 0 .
\end{equation}
Notably, numerical evidence shows us that the inequalities~\eqref{Eq.New1} and~\eqref{Eq.New2}  are less restrictive than Eq.~\eqref{Eq.Kravchuk}. 
\section{Shadow inequalities from the double-copy Hilbert space}
\label{App.Doublecopy}
Here, we derive the shadow inequalities~\eqref{Eq.Kravchuk} that define $\RR_2$ [see Eq.~\eqref{Eq.Range2}]. For that purpose, we use the double-copy of the initial Hilbert space of $N$ qubits, $\Hs_1^{\otimes N} \otimes \Hs_1^{\otimes N}$, where $\Hs_1$ is the Hilbert space of a single qubit. This double-copy technique has also been used to study the entanglement of mixed states in~\cite{PhysRevLett.95.260502,PhysRevLett.98.140505,PhysRevA.74.052303}. We start with a reformulation of the $S_{m}$ variables
\begin{equation}
    \begin{aligned}
        S_{m} = &  
        \sum_{\wt(\si_{\bm{\mu}})=m} |\bra{\psi} \si_{\bmu} \ket{\psi}|^2
        \\
        = 
        &
        \bra{\Psi} 
        \left( \sum_{\wt(\si_{\bm{\mu}})=m} \si_{\bmu} \otimes \si_{\bmu}
        \right)
        \ket{\Psi} 
        \\
        = & \bra{\Psi} \hat{S}_{m} \ket{\Psi} ,
    \end{aligned}
\end{equation}
where $\ket{\Psi}= \ket{\psi}\otimes \ket{\psi}$ and the sum runs over the multi-indices $\bmu$ with $m$ nonzero entries. Thus, the sector length $S_m$ of a pure state $\ket{\psi}$ can be thought of as the expectation value of an observable $\hat{S}_m$\footnote{We can also make this generalization for mixed states, $S_m(\rho) = \Tr [ (\rho \otimes \rho) \hat{S}_m]. $}.
The terms of the operators $\hat{S}_{m}$ are grouped as
\begin{equation}
\begin{aligned}    
&    \hat{S}_{m} = \sum_{\wt(\si_{\bm{\mu}})=m}  \si_{\bm{\mu}} \otimes  
    \si_{\bm{\mu}}
\\
& = \hspace{-0.2cm}  \sum_{\pi \in \mathrm{Sym}(N)} \sum_{a_1 , \dots a_m=1}^3 \hspace{-0.2cm}
    \frac{    \si_{\pi(a_1 ,\dots a_m \bm{0}_{N-m})} \otimes  
    \si_{\pi(a_1 ,\dots a_m \bm{0}_{N-m})}}{m! (N-m)!}
    ,
\end{aligned}
\end{equation}
where $\mathrm{Sym}(N)$ is the permutation group of $N$ elements, and the denominator accounts for multiplicities to ensure that each distinct term appears exactly once.
The index $a_{\alpha}$ appears twice, in the positions $\alpha$ and $N+\alpha$, and sums from 1 to 3. 
To calculate the eigenspectrum of $\hat{S}_{m}$, we start with the decomposition
\begin{equation}
\label{Eq.Dec.tensor.sigmas}
\begin{aligned}
\sum_{a=1}^3  & \big( \si_{a} \big)_{\alpha} \otimes  \big( \si_a \big)_{N+\alpha}
\\
={}&
\sum_{j=0}^1 \sum_{m=-j}^j 
    F(j) \ket{j,m}_{\alpha , N+\alpha} \bra{j,m}_{\alpha , N+\alpha}    
\\
={}& F(1) \PP^{(\alpha)}_1 + F(0) \PP^{(\alpha)}_{0} ,
\end{aligned}
\end{equation}
where 
\begin{equation}
\begin{aligned}    
F(j) \equiv{}& 2j(j+1) -3 = \left\{
\begin{array}{cc}
    1 & j=1 \\
    -3 & j=0 
\end{array}
\right.
,
\\
\PP^{(\alpha)}_1 \equiv &
\sum_{m=-1}^1
    \ket{1,m}_{\alpha , N+\alpha} \bra{1,m}_{\alpha , N+\alpha}   \, ,
\\
\PP^{(\alpha)}_0 \equiv &{}
\ket{0,0}_{\alpha , N+\alpha} \bra{0,0}_{\alpha , N+\alpha}  \, ,
\end{aligned}
\end{equation}
and $\ket{j,m}_{\alpha ,N+\alpha}$ denotes the corresponding triplet and singlet states of the two coupled qubits formed by the $\alpha$-th and $(N+\alpha)$-th qubits. 
 
Thus, the eigenvectors of $\hat{S}_{m}$ are
\begin{equation}
\label{Eq.Sk.Eigenspectrum}
\begin{aligned}    
&   \hat{S}_{m} \bigotimes_{\alpha=1}^N \ket{j_{\alpha} , m_{ \alpha} }_{\alpha,N+\alpha}
\\
={}& 
e_{m} (F(j_1) , \dots , F(j_N))
\bigotimes_{\alpha=1}^N \ket{j_{\alpha} , m_{\alpha} }_{\alpha,N+\alpha} ,
\end{aligned}
\end{equation}
where $e_{m}(x_1, \dots , x_N)$ is the $m$-th elementary symmetric function of $N$ variables
\begin{equation}
    e_{m} (x_1, \dots , x_N)
    \equiv
    \sum_{1 \leqslant j_1 < j_2 < \dots < j_{m} \leqslant N} x_{j_1} \dots x_{j_{m}} .
\end{equation}
We can observe that all the $\hat{S}_{m}$ operators have common eigenvectors. Consequently, $[\hat{S}_{m} , \hat{S}_{m'}]=0$. Also, each eigenvalue~\eqref{Eq.Sk.Eigenspectrum} has degeneracy $\tbinom{N}{g}3^{g}$ where $g$ is equal to the number of $j$ equal to 1. This implies that 
    \begin{equation}
    \label{Eq.SS.op.m}
        \hat{S}_{m} = \sum_{g=0}^N 
        \la_{m,g}
  \QQ_{g} , 
    \end{equation}
    with
    \begin{equation}
    \begin{aligned}        
        \la_{m,g}={}& e_{m} \Big( \underbrace{F(1),\dots ,F(1)}_{g} , \underbrace{F(0) , \dots , F(0)}_{N-g} \Big) .
\\
={}& e_{m} \Big( \underbrace{1,\dots ,1}_{g} , \underbrace{-3 , \dots , -3}_{N-g} \Big)
\\
={}&
\sum_{s=0}^{m} 
\mbinom{g}{s} \mbinom{N-g}{m-s} (-3)^{m-s} 
\\
={}&
(-1)^{m} K_{m}(g,N)
,
    \end{aligned}
    \end{equation}
where we have used the generating function of the symmetric polynomials~\cite{macdonald1998symmetric}, and the last expression are the Kravchuk polynomials~\eqref{Eq.Krav.poly}. The $\QQ_g$ operators are given by 
\begin{equation}
\label{Eq.QQ.g}
    \QQ_g =  \frac{1}{g! (N-g)!}
  \sum_{\pi \in \mathrm{Sym}(N)} \hspace{-0.2cm} 
    \Bigg( \bigotimes_{i=1}^g \PP^{\pi(i)}_1 \Bigg) \otimes 
    \Bigg( \bigotimes_{i=g+1}^{N} \PP^{\pi(i)}_0 \Bigg) .
\end{equation}
The state $\ket{\Psi}$ is symmetric under the swap operator $S_{12} (\ket{\psi_1} \otimes \ket{\psi_2} )= \ket{\psi_2} \otimes \ket{\psi_1}$. Over this permutation, $\ket{j,m_j}_{\alpha, N+\alpha}$ transforms to $(-1)^{j+1}\ket{j,m_j}_{\alpha, N+\alpha}$. Hence, $\QQ_{g} \ket{\Psi} = 0$ whenever $N-g$ is odd. 

By denoting the expectation value of the projections $\bra{ \Psi} \QQ_{g}  \ket{\Psi} = Q_{g} $, we can write $S_{m}$ as
\begin{equation}
     S_{m} = \sum_{g=0 }^N (-1)^{m} K_{m}(g,N) Q_g ,
\end{equation}
and $Q_g=0 $ for $N-g$ odd. Notably, the $Q_g$ coefficients are proportional to the shadow enumerators~\eqref{Eq.Inv.Sk}. To prove this, we use the orthogonality condition of the Kravchuk polynomials~\eqref{Eq.OR.Kravchuk} in the previous equation and in Eq.~\eqref{Eq.Inv.Sk}, which gives $S_g^{(e)} =2^N Q_g$. Therefore, $S_g^{(e)}=0 $ for $N-g$ odd. Additionally, since $\sum_{g=0}^N \QQ_g$ is equal to the identity matrix in the symmetric sector of $\Hs \otimes \Hs$, then 
\begin{equation}
\sum_{g=0}^{N} S_g^{(e)} = 2^N .   
\end{equation}
Consequently, there are only $\lfloor N/2 \rfloor$ linearly independent shadow enumerators.
\section{Multipartite concurrences}
\label{App.multipartite.concurrence}
Tensor products of the projectors $\PP^{(\alpha)}_x$ defined in Appendix~\ref{App.Doublecopy} are used in Refs.~\cite{PhysRevA.98.052317,PhysRevLett.95.260502,PhysRevLett.98.140505,PhysRevA.74.052303} to define multipartite concurrences\footnote{In Ref.~\cite{PhysRevA.74.052303}, the operator $\PP^{(\alpha)}_1$ ($\PP^{(\alpha)}_0$) is denoted by $P^{(\alpha)}_+$ ($P^{(\alpha)}_-$). }. 
For that, we define first a function $C_{\hat{\mathcal{A}}}(\ket{\psi})$ as
\begin{equation}
\label{Eq.Family.function}
    C_{\hat{\mathcal{A}}}(\ket{\psi}) = 2\sqrt{ \bra{\psi}\otimes \bra{\psi} \hat{\mathcal{A}} \ket{\psi}\otimes \ket{\psi} } ,
\end{equation}
where 
\begin{equation}
    \hat{\mathcal{A}} = \sum_{s_1, \dots , s_N=0}^1 p_{s_1 , \dots , s_N } \PP^{(1)}_{s_1} \otimes \dots \otimes \PP^{(N)}_{s_N} ,
\end{equation}
with the $p_{s_1 \dots s_N}$ coefficients such that
\begin{equation}
\label{Eq.pos.Asym}
\bra{\psi}\otimes \bra{\psi} \hat{\mathcal{A}} \ket{\psi} \otimes \ket{\psi} \geq 0 
\end{equation} 
for all $\ket{\psi} \in \Hs$~\cite{PhysRevA.74.052303}, \emph{i.e.}, $\hat{\mathcal{A}}$ is a positive semidefinite operator in the symmetric sector of $\Hs \otimes \Hs$\footnote{The positive semidefinite condition of $\hat{\mathcal{A}}$ over the symmetric sector of $\Hs \otimes \Hs$ holds, but not necessarily~\cite{PhysRevA.74.052303}, when all the $p_{s_1 , \dots , s_N}$  coefficients are positive.}. For example, the $\hat{S}_m$ (Eq.~\eqref{Eq.SS.op.m}) and $\QQ_g$ (Eq.~\eqref{Eq.QQ.g}) operators fulfill~\eqref{Eq.pos.Asym} and, consequently, they belong to this family of functions. In particular, if $C_{\hat{\mathcal{A}}}(\ket{\psi})$ is an entanglement monotone, then it qualifies as a well-defined multipartite concurrence~\cite{PhysRevA.74.052303,PhysRevA.98.052317}. References~\cite{PhysRevA.74.052303,PhysRevA.98.052317} discuss necessary and/or sufficient conditions for $C_{\hat{\mathcal{A}}}$ to be a multipartite concurrence. In particular, \cite{PhysRevA.98.052317} shows that the quantity $R_{\rho}$ in Eq.~\eqref{psipsipsitilde}, which is equal to $\Shpsi_0 (\rho)$~\eqref{Eq.Shadow.Rrho}  for pure states, is a concurrence. The latter result  has been discussed previously in other references~\cite{PhysRevA.63.044301,10.1063/1.1723701,PhysRevA.85.022301}. 

The interpretation of the shadow inequalities as the expectation value of a projector in the double-copy Hilbert space was briefly discussed in Ref.~\cite{PhysRevA.98.052317}\footnote{Specifically, in Eqs.~(14) and (18) as well as footnote [45] in Ref.~\cite{PhysRevA.98.052317}.}. Finally, note that the concurrences can also be generalized to mixed states via the convex-roof extension~\cite{PhysRevA.74.052303}. 
\section{Optimal states}
\label{App.states.table}
In this Appendix, we present analytical expressions for proven or putative optimal states that extremize certain sector lengths or $\Tr (\rho_A^2) + R_{\rho_A}$. Most of these states have been deduced from numerical optimization. Their extremality is then either rigorously established based on our inequalities and those existing in the literature, or supported by numerical optimization, see Tables~\ref{tab.3} and \ref{tab.4} for more details.

Before presenting the states, we recall that a $k$-uniform $N$-qubit state is one for which all the reduced density matrices of at most $k$ qubits are maximally mixed. When $k=\lfloor N/2 \rfloor $, such states are known as AME$(N,2)$. Consequently, for a $k$-uniform state, all sector lengths $S_m$ vanish for $m\leq k$; that is, $S_m = 0$ for all $m \leq k$.

As some optimal states are symmetric, it is useful to express them in terms of symmetric $N$-qubit Dicke states
\begin{equation}
\label{Eq.Dicke}
\ket{D^{(\alpha)}_N} = \mbinom{N}{\alpha}^{-1/2} \sum_{\pi}  \big| \pi(\underbrace{00 \dots 0}_{N-\alpha} \underbrace{11 \dots 1}_{\alpha} )\big\rangle,
\end{equation}  
where $\alpha = 0, \dots, N$, and the sum runs over all permutations $\pi$ of the $N$ qubit labels. Symmetric $N$-qubit states can be uniquely represented by configurations of Majorana points on the Bloch sphere, known as Majorana stellar constellation~\cite{Majorana_1932,Zim:06}. In the following, certain state names explicitly refer to this geometric representation.

\subsection*{\texorpdfstring{$N=4$}{Lg}} 
$\bullet$ The symmetric $4$-qubit tetrahedron state, given by  
\begin{equation}
\begin{aligned}    
\label{tetrapsi}
\ket{\textrm{tetra}} = \frac{1}{2} \left( \ket{D_4^{(0)}} + i \sqrt{2} \ket{D_4^{(2)}} + \ket{D_4^{(4)}} \right),
\end{aligned}
\end{equation}  
minimizes both $S_1$ and $S_2$ and maximizes $S_3$. It has sector lengths
\begin{equation}
\mathbf{S}= \left(0,2,8,5\right).
\end{equation}
Among all symmetric four-qubit states, it stands out as the unique maximally entangled one~\cite{Aulbach2010,Martin2010,PhysRevA.90.032314}. Each of its two-qubit reduced density matrices is maximally mixed within the symmetric subspace, having nonzero eigenvalues $(\tfrac{1}{3},\tfrac{1}{3},\tfrac{1}{3})$. Its extremality has been established analytically.

\subsection*{\texorpdfstring{$N=5$}{Lg}} 
$\bullet$ The $2$-uniform $5$-qubit state~\cite{PhysRevA.86.052335,Borras_2007,AME.Huber}
\begin{equation}
\begin{aligned}    
    \label{Eq.AME52.def}
    \ket{\mathrm{AME}(5,2)}={}& \tfrac{1}{\sqrt{8}} \big( 
      |00000\rangle 
    + |00011\rangle 
    + |01101\rangle 
    + |01110\rangle 
    \\ &
    + |10101\rangle 
    - |10110\rangle 
    + |11000\rangle 
    - |11011\rangle
    \big)
\end{aligned}
\end{equation}
minimizes both $S_1$ and $S_2$ and maximizes $S_4$. It has sector lengths
\begin{equation}
\mathbf{S}= \left(0,0,10,15,6\right).
\end{equation}
All extremality properties have been established analytically.\\

$\bullet$ The state
\begin{equation}\label{psi_5}
\begin{aligned}    
    \ket{\psi_5}={}& \tfrac{1}{\sqrt{8}} \big( 
      |00000\rangle 
    + |00011\rangle 
    - |00101\rangle 
    + |00110\rangle 
    \\ &
    + |10000\rangle 
    + |10011\rangle 
    + |10101\rangle 
    - |10110\rangle
    \big)
\end{aligned}
\end{equation}
has sector lengths
\begin{equation}
\mathbf{S}= \left(1,2,10,13,5\right)
\end{equation}
and is used in the superposition \eqref{superposition_psi5} to cover part of the right boundary of $\RR$.

\subsection*{\texorpdfstring{$N=6$}{Lg}} 
$\bullet$ The $2$-uniform $6$-qubit state~\cite{PhysRevA.86.052335,Borras_2007,AME.Huber}
\begin{equation}\label{Eq.AME62.def}
    \begin{aligned}        
 &   \ket{\mathrm{AME}(6,2)}= \tfrac{1}{5}\big( 
    |000000\rangle 
    + |000011\rangle 
    + |000101\rangle  
\\
&\quad
    + |000110\rangle 
    + |001001\rangle  
    - |001010\rangle 
    - |001100\rangle 
 \\
&\quad   + |001111\rangle 
    + |010001\rangle 
    + |010100\rangle 
    - |010111\rangle 
\\
&\quad   + |011000\rangle 
    - |011011\rangle  
    + |011101\rangle 
    - |011111\rangle 
\\  
&\quad
    - |100001\rangle 
    - |100100\rangle 
    + |100111\rangle 
    + |101000\rangle 
\\
&\quad   + |101011\rangle 
    + |101110\rangle 
    + |110000\rangle 
    + |110011\rangle 
\\
&\quad
    + 
  |110101\rangle + 
  |111000\rangle \big)
    \end{aligned}
\end{equation}
minimizes both $S_1$ and $S_2$ and maximizes $S_4$. It has sector lengths
\begin{equation}
\mathbf{S}= \left(0,0,0,45,0,18\right).
\end{equation}
Its extremal properties have been established analytically.\\[5pt]

$\bullet$ The $2$-uniform $6$-qubit state
\begin{equation}
    \begin{aligned}
\label{Eq.psi6}
|\psi_6\rangle ={}& \tfrac{1}{4} \big(
 |000000\rangle + |000111\rangle + |001110\rangle + |010101\rangle \\
& + |100100\rangle + |110001\rangle + |110110\rangle + |111111\rangle
  \\
 &-|001001\rangle - |010010\rangle - |011011\rangle - |011100\rangle \\
& - |100011\rangle - |101010\rangle - |101101\rangle - |111000\rangle
\big)      
    \end{aligned}
\end{equation}
minimizes both $S_1$ and $S_2$. It has $12$ of its $3$-qubit reduced density matrices maximally mixed. The remaining $8$ have an identical spectrum $(0,0,0,0,\tfrac{1}{4},\tfrac{1}{4},\tfrac{1}{4},\tfrac{1}{4})$. The sector lengths for this state are
\begin{equation}
\mathbf{S}= \left(0,0,8,21,24,10\right),
\end{equation}
corresponding to a vertex of the polytope shown in Fig.~\ref{fig3}.\\[5pt]

$\bullet$ The $1$-uniform $6$-qubit symmetric state\footnote{Its name comes from its Majorana constellation consisting of a square pyramid (the top vertex is twice degenerate) whose base is at a polar angle $\theta=-2 \arctan (3^{1/4} 5^{1/8})$.}
\begin{equation}
\label{pyramidN6}    |\mathrm{pyramid}\rangle=-\tfrac{1}{2}|D_6^{(0)}\rangle+\tfrac{\sqrt{3}}{2}|D_6^{(4)}\rangle
\end{equation}
minimizes $S_1$ and maximizes $S_5$. It has sector lengths
\begin{equation}
\mathbf{S}= \left(0,\frac{27}{5},8,\frac{51}{5},24,\frac{77}{5}\right) .
\end{equation}
Its extremal properties have been established analytically.

\begin{widetext}
\subsection*{\texorpdfstring{$N=7$}{Lg}}
$\bullet$ The $1$-uniform $7$-qubit symmetric state
\begin{equation}\label{psi7}
|\psi_7\rangle=\mathcal{N}\left(2\sqrt{2}\,|D_7^{(0)}\rangle+\sqrt{14}\,|D_7^{(3)}\rangle+\sqrt{14}\,|D_7^{(6)}\rangle\right)
\end{equation}
minimizes both $S_1$ and $S_4$ and maximizes $S_6$. Its sector lengths are given by
\begin{equation}
\mathbf{S}= \left(0,7,14,7,28,49,22\right) .
\end{equation}
All its two-qubit reduced states have nonzero eigenvalues $(\tfrac{1}{3}, \tfrac{1}{3}, \tfrac{1}{3})$, while its three-qubit reduced states have nonzero eigenvalues $(\tfrac{1}{10}, \tfrac{2}{10}, \tfrac{3}{10}, \tfrac{4}{10})$. Its extremality for $S_4$ and $S_6$ has not been established analytically but is supported by our numerical results.\\[5pt]

$\bullet$ The $2$-uniform $7$-qubit state given in Eq.~(17) of Ref.~\cite{Zha_2012}, which reads
\begin{equation}\label{psi1234567}
\begin{aligned}
|\psi_M\rangle_{1234567} = \tfrac{1}{4\sqrt{2}} \Big[{}&
\left|0000000\right\rangle + \left|0000011\right\rangle + \left|0001101\right\rangle + \left|0001110\right\rangle 
+ \left|0010001\right\rangle
- \left|0010010\right\rangle + \left|0011100\right\rangle - \left|0011111\right\rangle
\\ & - \left|0100101\right\rangle - \left|0100110\right\rangle + \left|0101000\right\rangle + \left|0101011\right\rangle 
+ \left|0110100\right\rangle - \left|0110111\right\rangle - \left|0111001\right\rangle + \left|0111010\right\rangle 
\\[3pt] & - \left|1000100\right\rangle - \left|1000111\right\rangle + \left|1001001\right\rangle + \left|1001010\right\rangle 
+ \left|1010101\right\rangle - \left|1010110\right\rangle - \left|1011000\right\rangle + \left|1011011\right\rangle \\ & 
+ \left|1100001\right\rangle + \left|1100010\right\rangle + \left|1101100\right\rangle + \left|1101111\right\rangle 
+ \left|1110000\right\rangle - \left|1110011\right\rangle + \left|1111101\right\rangle - \left|1111110\right\rangle \Big].
\end{aligned}
\end{equation}
minimizes both $S_1$ and $S_2$, which are both zero. This state has $32$ of its $3$-qubit reduced density matrices maximally mixed. The remaining three have the spectrum $(0, 0, 0, 0, \tfrac{1}{4}, \tfrac{1}{4}, \tfrac{1}{4}, \tfrac{1}{4})$. Its sector lengths are
\begin{equation}
\mathbf{S}= \left( 0,0,3,29,42,34,19 \right) .
\end{equation}

$\bullet$ The $7$-qubit state 
\begin{equation}
\label{Eq.psi.num}
\begin{aligned}
|\psi_{7b}\rangle = & \frac{1}{\sqrt{13}} \Big[
\sqrt{\tfrac{3}{2}} \ket{0000}\otimes\big(  \ket{001}
-\ket{100}\big)
+ \ket{0001010}
+ \ket{0101000}
+ \ket{0110000}
+ \ket{0011000}\\
&\quad
+ \omega \ket{1100000}
+ \omega^2 \ket{0010010}
+ \omega^3 \ket{1010000}
+ \omega^4 \ket{0100010}
+ \omega^5 \ket{100}\otimes\big(\ket{0010}
+\ket{1000}\big)
\Big],
\end{aligned}
\end{equation}
with $\omega=e^{i\pi/3}$, maximizes $S_5$ and has sector lengths
\begin{equation}
\mathbf{S}= \left( \frac{25}{13}, \frac{9}{13}, \frac{125}{13}, 35, \frac{603}{13}, \frac{355}{13}, \frac{79}{13} \right) .
\end{equation}
Its extremality has not been established analytically but is supported by our numerics.
\subsection*{\texorpdfstring{$N=8$}{Lg}} 
$\bullet$ The $3$-uniform $8$-qubit state given in Eq.~(19) of Ref.~\cite{Zha_2013}, which reads
\begin{equation}\label{psi12783456}
\begin{aligned}
|\psi_M\rangle_{12783456}
= \frac{1}{8} \Big[&{}
\Big( \left|0000\right\rangle + \left|0011\right\rangle - \left|1101\right\rangle + \left|1110\right\rangle \Big)_{1278}
\otimes \Big( \left|0000\right\rangle + \left|0111\right\rangle - \left|1001\right\rangle + \left|1110\right\rangle \Big)_{3456} \\
& + \Big(-\left|0001\right\rangle + \left|0010\right\rangle + \left|1100\right\rangle + \left|1111\right\rangle \Big)_{1278}
\otimes \Big( \left|0001\right\rangle + \left|0110\right\rangle + \left|1000\right\rangle - \left|1111\right\rangle \Big)_{3456} \\
& + \Big( \left|0100\right\rangle - \left|0111\right\rangle + \left|1001\right\rangle + \left|1010\right\rangle \Big)_{1278}
\otimes \Big( -\left|0011\right\rangle + \left|0100\right\rangle + \left|1010\right\rangle + \left|1101\right\rangle \Big)_{3456} \\
& + \Big( \left|0101\right\rangle + \left|0110\right\rangle + \left|1000\right\rangle - \left|1011\right\rangle \Big)_{1278}
\otimes \Big( -\left|0010\right\rangle + \left|0101\right\rangle - \left|1011\right\rangle - \left|1100\right\rangle \Big)_{3456} \Big] ,
\end{aligned}
\end{equation}
where the multi-indices denote the qubit labels, minimizes both $S_1, S_2$ and $S_3$, which are all zero. Its sector lengths are given by
\begin{equation}
\mathbf{S} = (0,0,0,26,64,72,64,29) .
\end{equation}

$\bullet$ The $1$-uniform $8$-qubit symmetric state 
\begin{equation}\label{tetra8}
    |\mathrm{tetra}(8)\rangle=\frac{1}{3\sqrt{3}}\left(\sqrt{7}\,|D_8^{(0)}\rangle+2\,|D_8^{(3)}\rangle+4\,|D_8^{(6)}\rangle\right)
\end{equation}
minimizes $S_1$ and maximizes $S_7$. This state has doubly degenerate Majorana stars at the vertices of a tetrahedron. Its sector lengths are given by
\begin{equation}
\mathbf{S}= \left( 0,\frac{28}{3},\frac{144}{7},\frac{310}{21},\frac{160}{7},\frac{1396}{21},\frac{592}{7},\frac{255}{7}\right) .
\end{equation}
Its extremality for $S_7$ has not been established analytically but is supported by our numerics.

$\bullet$ The $8$-qubit state $|\mathrm{tetra}\rangle^{\otimes 2}$ with $|\mathrm{tetra}\rangle$ given in Eq.~\eqref{tetrapsi} minimizes $S_1$ and $S_4$ and has sector lengths
\begin{equation}
\mathbf{S}= \left(0,4,16,14,32,84,80,25\right) .
\end{equation}
Its extremality for $S_4$ has not been established analytically but is supported by our numerics.

$\bullet$ The 3-uniform $8$-qubit state given in Eq.~(12) of Ref.~\cite{Zha2018}, which reads
\begin{equation}\label{psi8_3uniform}
\begin{aligned}
|\psi^1_{12345678}\rangle = \tfrac{1}{4\sqrt{2}} \Big[
   & \Big(|0000\rangle + |1111\rangle \Big)_{1256} \otimes  \Big(|0000\rangle + |0011\rangle + |1100\rangle + |1111\rangle \Big)_{3478} \\
    & +  \Big(|0011\rangle + |1100\rangle \Big)_{1256} \otimes \Big(|0110\rangle + |0101\rangle + |1010\rangle + |1001\rangle \Big)_{3478} \\
    & +  \Big(|0101\rangle + |1010\rangle \Big)_{1256} \otimes \Big(|0110\rangle - |0101\rangle - |1010\rangle + |1001\rangle \Big)_{3478} \\
    & +  \Big(|0110\rangle + |1001\rangle \Big)_{1256} \otimes \Big(|0000\rangle - |0011\rangle - |1100\rangle + |1111\rangle \Big)_{3478}
\Big].
\end{aligned}
\end{equation}
minimizes $S_1, S_2, S_3, S_5$ and $S_7$ and maximizes $S_6$. Its sector lengths are given by
\begin{equation}
\label{Eq.psi.num.prime}
\mathbf{S} = (0,0,0,42,0,168,0,45) .
\end{equation}
Its extremality for $S_6$ has not been established analytically but is supported by our numerics.

\subsection*{\texorpdfstring{$N=9$}{Lg}} 
$\bullet$ The $3$-uniform $9$-qubit state given in Eq.~(15) of Ref.~\cite{Che2020}, which reads
\begin{equation}
\label{psi9qubits}
\begin{aligned}
|\psi_{m}\rangle_{123456789} = \frac{1}{\sqrt{32}}\big[&
|000000000\rangle
- |000000111\rangle
- |000011001\rangle
+ |000011110\rangle
+ |001101001\rangle
+ |001101110\rangle\\
& + |001110000\rangle
+ |001110111\rangle
 + |010101011\rangle
- |010101100\rangle
- |010110010\rangle
+ |010110101\rangle \\
& + |011000010\rangle
+ |011000101\rangle
 + |011011011\rangle
+ |011011100\rangle
 - |100100011\rangle
- |100100100\rangle\\
& - |100111010\rangle
- |100111101\rangle
- |101001010\rangle
 + |101001101\rangle
+ |101010011\rangle
- |101010100\rangle\\
& 
+ |110001000\rangle
 + |110001111\rangle
+ |110010001\rangle
+ |110010110\rangle
 + |111100001\rangle
- |111100110\rangle\\
& - |111111000\rangle
+ |111111111\rangle
\big]
\end{aligned}
\end{equation}
minimizes $S_1, S_2, S_3$. Its sector lengths are given by
\begin{equation}
\label{Eq.9qubits}
\mathbf{S} = (0,0,0,18,72,120,144,117,40) .
\end{equation}

\renewcommand{\arraystretch}{1.3} 
\setlength{\arrayrulewidth}{0.5pt} 

    \begin{table*}[t]
\mbox{
\begin{minipage}[t]{0.4\textwidth}
\vspace{0pt}
\begin{tabular}{|c|@{\hspace{4pt}}c@{\hspace{4pt}}c|@{\hspace{4pt}}c@{\hspace{4pt}}c|}
\hline
$\phantom{\Big|}\;N\;\phantom{\Big|}$ & $\min S_m$ & Optimal state & $\max S_m$ & Optimal state 
\\
\hline \hline
\multicolumn{5}{|c|}{$m = 1$} 
\\
\hline
$N \geq 2$ & 0\footnote{\hypertarget{FOOT}{$\phantom{a}$} \hspace{-0.4cm} $S_m \geq 0$ 
\hspace{0.1cm} 
\textsuperscript{b} \cite{wyderka2020characterizing} \hspace{0.1cm} 
\textsuperscript{c} \cite{PhysRevA.94.042302} 
\hspace{0.1cm}
\textsuperscript{d} \cite{HIGUCHI2000213,PhysRevA.69.052330}
\hspace{0.1cm}
\textsuperscript{e} \cite{PhysRevA.86.052335,Borras_2007,AME.Huber} 
\hspace{0.1cm}
\textsuperscript{f} \cite{Eltschka2020}}
& \bcolor $|\mathrm{GHZ}(N)\rangle$ & $N$\textsuperscript{\hyperlink{FOOT}{b}} & \bcolor $|0\rangle^{\otimes N}$ 
\\
\hline \hline 
\multicolumn{5}{|c|}{$m = 2$} 
\\
\hline
2 & 1\textsuperscript{\hyperlink{FOOT}{c}} & \bcolor $|0\rangle^{\otimes 2}$ & 3\textsuperscript{\hyperlink{FOOT}{b}} & \bcolor $|\mathrm{GHZ}(2)\rangle$ 
\\
3 & 3\textsuperscript{\hyperlink{FOOT}{b}} & \bcolor any $|\psi\rangle$ & 3\textsuperscript{\hyperlink{FOOT}{b}} & \bcolor any $|\mathrm{\psi}\rangle$ 
\\
4 & 2\textsuperscript{\hyperlink{FOOT}{d}} & \bcolor $|\mathrm{tetra}\rangle$ & 6\textsuperscript{\hyperlink{FOOT}{b}} & \bcolor  $|0\rangle^{\otimes 4}$ 
\\
5 & 0\textsuperscript{\hyperlink{FOOT}{e}} & \bcolor $|\mathrm{AME}(5,2)\rangle^\eqref{Eq.AME52.def}$ & 10\textsuperscript{\hyperlink{FOOT}{b}} & \bcolor $|0\rangle^{\otimes 5}$ 
\\
6 & 0\textsuperscript{\hyperlink{FOOT}{e}} & \bcolor $|\mathrm{AME}(6,2)\rangle^\eqref{Eq.AME62.def}$ & 15\textsuperscript{\hyperlink{FOOT}{b}} & \bcolor $|0\rangle^{\otimes 6}$ 
\\
7 & 0\textsuperscript{\hyperlink{FOOT}{b}} & \bcolor $|\mathrm{AME(6,2)}\rangle |0\rangle$ & 21\textsuperscript{\hyperlink{FOOT}{b}} & \bcolor $|0\rangle^{\otimes 7}$ 
\\
8 & 0$\phantom{^a}$ & \bcolor $|\psi_M\rangle_{12783456}^{\eqref{psi12783456}}$ & 28\textsuperscript{\hyperlink{FOOT}{b}} & \bcolor $|0\rangle^{\otimes 8}$ 
\\
$N \geq 9$ & 0$\phantom{^a}$ & \bcolor see Sec.~\ref{Subsec.1.2.body}  & $\mbinom{N}{2}$\textsuperscript{\hyperlink{FOOT}{b}} & \bcolor $|0\rangle^{\otimes N}$
\\[0.1cm]
\hline \hline
\multicolumn{5}{|c|}{$m = 3$} 
\\
\hline
3 & 1\textsuperscript{\hyperlink{FOOT}{c}} & \bcolor $|0\rangle^{\otimes 3}$ & 4\textsuperscript{\hyperlink{FOOT}{c}} & \bcolor $|\mathrm{GHZ}(3)\rangle$ 
\\
4 & 0\textsuperscript{\hyperlink{FOOT}{a}} & \bcolor $|\mathrm{GHZ}(4)\rangle$ & 8\textsuperscript{\hyperlink{FOOT}{b}} &  \bcolor $|\mathrm{tetra}\rangle$ 
\\
$N \geq 5$ & 0\textsuperscript{\hyperlink{FOOT}{a}} & \bcolor $|\mathrm{GHZ}(N)\rangle$ & $\mbinom{N}{3}$\textsuperscript{\hyperlink{FOOT}{b}} & \bcolor $|0\rangle^{\otimes N}$ 
\\[0.1cm]
\hline \hline
\multicolumn{5}{|c|}{$m = 4$} 
\\
\hline
4 & 1\textsuperscript{\hyperlink{FOOT}{c}} & \bcolor $|0\rangle^{\otimes 4}$ & 9\textsuperscript{\hyperlink{FOOT}{b}} & \bcolor $|\mathrm{GHZ}(4)\rangle$ 
\\
5 & 5$\phantom{^a}$ & \bcolor $|0\rangle^{\otimes 5}$ & 15$\phantom{^a}$ & \bcolor $|\mathrm{AME}(5,2)\rangle$ 
\\
6 & 5$\phantom{^a}$ & $|\mathrm{GHZ}(5)\rangle |0\rangle$ & 45$\phantom{^a}$ & \bcolor $|\mathrm{AME}(6,2)\rangle$ 
\\
7 & 7$\phantom{^a}$ & $|\psi_7\rangle^{\eqref{psi7}}$ & 45$\phantom{^a}$ & $|\mathrm{AME(6,2)}\rangle |0\rangle$ 
\\
8 & 14$\phantom{^a}$ & $|\mathrm{tetra}\rangle^{\otimes 2}$ & 70$\phantom{^a}$ & $|0\rangle^{\otimes 8}$ 
\\
\hline \hline
\multicolumn{5}{|c|}{$m = 5$} 
\\
\hline
5 & 1\textsuperscript{\hyperlink{FOOT}{c}} & \bcolor $|0\rangle^{\otimes 5}$ & 16\textsuperscript{\hyperlink{FOOT}{c}} & \bcolor $|\mathrm{GHZ}(5)\rangle$ 
\\
6 & 0\textsuperscript{\hyperlink{FOOT}{a}} & \bcolor $|\mathrm{GHZ}(6)\rangle$ & 24$\phantom{^a}$ & \bcolor $|\mathrm{pyramid}\rangle^{\eqref{pyramidN6}}$ 
\\
7 & 0\textsuperscript{\hyperlink{FOOT}{a}} & \bcolor $|\mathrm{GHZ}(7)\rangle$ & 603/13$\phantom{^a}$ & $|\psi_{7b}\rangle^{\eqref{Eq.psi.num}}$
\\
8 & 0\textsuperscript{\hyperlink{FOOT}{a}} & \bcolor $|\mathrm{GHZ}(8)\rangle$ & 90$\phantom{^a}$ & $|\mathrm{AME(6,2)}\rangle |0\rangle^{\otimes 2}$ 
\\
$N \geq 9$ & 0\textsuperscript{\hyperlink{FOOT}{a}} & \bcolor $|\mathrm{GHZ}(N)\rangle$ & ? & ? 
\\
\hline \hline
\multicolumn{5}{|c|}{$m = 6$} 
\\
\hline
6 & 1\textsuperscript{\hyperlink{FOOT}{c}} & \bcolor $|0\rangle^{\otimes 6}$ & 33\textsuperscript{\hyperlink{FOOT}{b}} & \bcolor $|\mathrm{GHZ}(6)\rangle$ 
\\
7 & 7$\phantom{^a}$ & $|0\rangle^{\otimes 7}$ & 49$\phantom{^a}$ &  $|\psi_7\rangle^{\eqref{psi7}}$ \\
8 & 7$\phantom{^a}$ & $|\mathrm{GHZ}(7)\rangle |0\rangle$ & 168$\phantom{^a}$ & $|\psi^1_{12345678}\rangle^{\eqref{psi8_3uniform}}$ 
\\
\hline \hline
\multicolumn{5}{|c|}{$m = 7$} 
\\
\hline
7 & 1\textsuperscript{\hyperlink{FOOT}{c}} & \bcolor $|0\rangle^{\otimes 7}$ & 64\textsuperscript{\hyperlink{FOOT}{c}} & \bcolor $|\mathrm{GHZ}(7)\rangle$ 
\\
8 & 0\textsuperscript{\hyperlink{FOOT}{a}} & \bcolor $|\mathrm{GHZ}(8)\rangle$ & 592/7$\phantom{^a}$ & $|\mathrm{tetra}(8)\rangle^{\eqref{tetra8}}$ \\
$N \geq 9$ & 0\textsuperscript{\hyperlink{FOOT}{a}} & \bcolor $|\mathrm{GHZ}(N)\rangle$ & ? &  ?
\\
\hline \hline
\multicolumn{5}{|c|}{$m = 8$} \\
\hline
8 & 1\textsuperscript{\hyperlink{FOOT}{c}} & \bcolor $|0\rangle^{\otimes 8}$ & 129\textsuperscript{\hyperlink{FOOT}{b}} & \bcolor $|\mathrm{GHZ}(8)\rangle$
\\
\hline \hline
\multicolumn{5}{|c|}{$m = N$} \\
\hline
$N$ odd & 1\textsuperscript{\hyperlink{FOOT}{c}} & \bcolor $|0\rangle^{\otimes N}$ & $2^{N-1}$\textsuperscript{\hyperlink{FOOT}{c}} & \bcolor $|\mathrm{GHZ}(N)\rangle$
\\
\hline 
$N$ even & 1\textsuperscript{\hyperlink{FOOT}{c}} & \bcolor $|0\rangle^{\otimes N}$ & $2^{N-1}+1\textsuperscript{\hyperlink{FOOT}{f}}$ & \bcolor $|\mathrm{GHZ}(N)\rangle$
\\
\hline
\end{tabular}
\end{minipage}
\begin{minipage}[t]{0.7\textwidth}
\vspace{0pt}
\begin{tabular}{|c|@{\hspace{4pt}}c@{\hspace{4pt}}c|@{\hspace{4pt}}c@{\hspace{4pt}}c|}
\hline
$\phantom{\Big|}\;m\;\phantom{\Big|}$ & $\min S_m$ & optimal state & $\max S_m$ & optimal state
\\
\hline
\hline
\multicolumn{5}{|c|}{$N = 2$} \\
\hline
1 & 0\textsuperscript{\hyperlink{FOOT}{a}} 
& \bcolor $|\mathrm{GHZ}(2)\rangle$  & 2\textsuperscript{\hyperlink{FOOT}{b}} & \bcolor $|0\rangle^{\otimes 2}$ 
\\
2 & 1\textsuperscript{\hyperlink{FOOT}{c}} & \bcolor $|0\rangle^{\otimes 2}$ & 3\textsuperscript{\hyperlink{FOOT}{b}} & \bcolor $|\mathrm{GHZ}(2)\rangle$ 
\\
\hline \hline
\multicolumn{5}{|c|}{$N = 3$} \\
\hline
1 & 0\textsuperscript{\hyperlink{FOOT}{a}} & \bcolor $|\mathrm{GHZ}(3)\rangle$ & 3\textsuperscript{\hyperlink{FOOT}{b}} & \bcolor $|0\rangle^{\otimes 3}$ 
\\
2 & 3\textsuperscript{\hyperlink{FOOT}{b}} & \bcolor any $|\psi\rangle$ & 3\textsuperscript{\hyperlink{FOOT}{b}} & \bcolor any $|\psi\rangle$ 
\\
3 & 1\textsuperscript{\hyperlink{FOOT}{c}} & \bcolor $|0\rangle^{\otimes 3}$ & 4\textsuperscript{\hyperlink{FOOT}{c}} & \bcolor $|\mathrm{GHZ}(3)\rangle$ 
\\
\hline \hline
\multicolumn{5}{|c|}{$N = 4$} 
\\
\hline
1 & 0\textsuperscript{\hyperlink{FOOT}{a}} & \bcolor $|\mathrm{GHZ}(4)\rangle$ & 4\textsuperscript{\hyperlink{FOOT}{b}} & \bcolor $|0\rangle^{\otimes 4}$ 
\\
2 & 2\textsuperscript{\hyperlink{FOOT}{d}} & \bcolor $|\mathrm{tetra}\rangle$ & 6\textsuperscript{\hyperlink{FOOT}{b}} & \bcolor $|0\rangle^{\otimes 4}$ 
\\
3 & 0\textsuperscript{\hyperlink{FOOT}{a}} & \bcolor $|\mathrm{GHZ}(4)\rangle$ & 8\textsuperscript{\hyperlink{FOOT}{b}} & \bcolor $|\mathrm{tetra}\rangle$ 
\\
4 & 1\textsuperscript{\hyperlink{FOOT}{c}} & \bcolor $|0\rangle^{\otimes 4}$ & 9\textsuperscript{\hyperlink{FOOT}{b}} & \bcolor $|\mathrm{GHZ}(4)\rangle$ 
\\
\hline \hline
\multicolumn{5}{|c|}{$N = 5$} 
\\
\hline
1 & 0\textsuperscript{\hyperlink{FOOT}{a}} & \bcolor $|\mathrm{GHZ}(5)\rangle$ & 5\textsuperscript{\hyperlink{FOOT}{b}} & \bcolor $|0\rangle^{\otimes 5}$ 
\\
2 & 0\textsuperscript{\hyperlink{FOOT}{e}} & \bcolor $|\mathrm{AME}(5,2)\rangle$ & 10\textsuperscript{\hyperlink{FOOT}{b}} & \bcolor $|0\rangle^{\otimes 5}$ 
\\
3 & 0\textsuperscript{\hyperlink{FOOT}{a}} & \bcolor $|\mathrm{GHZ}(5)\rangle$ & 10\textsuperscript{\hyperlink{FOOT}{b}} & \bcolor $|0\rangle^{\otimes 5}$ 
\\
4 & 5$\phantom{^a}$ & \bcolor $|0\rangle^{\otimes 5}$ & 15$\phantom{^a}$ & \bcolor $|\mathrm{AME}(5,2)\rangle$ 
\\
5 & 1\textsuperscript{\hyperlink{FOOT}{c}} & \bcolor $|0\rangle^{\otimes 5}$ & 16\textsuperscript{\hyperlink{FOOT}{c}} & \bcolor $|\mathrm{GHZ}(5)\rangle$ 
\\
\hline \hline
\multicolumn{5}{|c|}{$N = 6$} 
\\
\hline
1 & 0\textsuperscript{\hyperlink{FOOT}{a}} & \bcolor $|\mathrm{GHZ}(6)\rangle$ & 6\textsuperscript{\hyperlink{FOOT}{b}} & \bcolor$|0\rangle^{\otimes 6}$ 
\\
2 & 0\textsuperscript{\hyperlink{FOOT}{e}} & \bcolor $|\mathrm{AME}(6,2)\rangle$ & 15\textsuperscript{\hyperlink{FOOT}{b}} & \bcolor $|0\rangle^{\otimes 6}$ 
\\
3 & 0\textsuperscript{\hyperlink{FOOT}{a}} & \bcolor $|\mathrm{GHZ}(6)\rangle$ & 20\textsuperscript{\hyperlink{FOOT}{b}} & \bcolor $|0\rangle^{\otimes 6}$ 
\\
4 & 5$\phantom{^a}$ &  $|\mathrm{GHZ}(5)\rangle|0\rangle$ & 45$\phantom{^a}$ & \bcolor $|\mathrm{AME}(6,2)\rangle$ 
\\
5 & 0\textsuperscript{\hyperlink{FOOT}{a}} & \bcolor $|\mathrm{GHZ}(6)\rangle$ & 24$\phantom{^a}$ & \bcolor $|\mathrm{pyramid}\rangle^{\eqref{pyramidN6}}$ 
\\
6 & 1\textsuperscript{\hyperlink{FOOT}{c}} & \bcolor $|0\rangle^{\otimes 6}$ & 33\textsuperscript{\hyperlink{FOOT}{b}} & \bcolor $|\mathrm{GHZ}(6)\rangle$ 
\\
\hline \hline
\multicolumn{5}{|c|}{$N = 7$} 
\\
\hline
1 & 0\textsuperscript{\hyperlink{FOOT}{a}} & \bcolor $|\mathrm{GHZ}(7)\rangle$ & 7\textsuperscript{\hyperlink{FOOT}{b}} & \bcolor $|0\rangle^{\otimes 7}$ 
\\
2 & 0$\phantom{^a}$ & \bcolor $|\mathrm{AME(6,2)}\rangle|0\rangle$ & 21\textsuperscript{\hyperlink{FOOT}{b}} & \bcolor $|0\rangle^{\otimes 7}$ 
\\
3 & 0\textsuperscript{\hyperlink{FOOT}{a}} & \bcolor $|\mathrm{GHZ}(7)\rangle$ & 35\textsuperscript{\hyperlink{FOOT}{b}} & \bcolor $|0\rangle^{\otimes 7}$ 
\\
4 & 7$\phantom{^a}$ & $|\psi_7\rangle^{\eqref{psi7}}$ & 45$\phantom{^a}$ & $|\mathrm{AME(6,2)}\rangle |0\rangle$ 
\\
5 & 0\textsuperscript{\hyperlink{FOOT}{a}} & \bcolor $|\mathrm{GHZ}(7)\rangle$ & 603/13$\phantom{^a}$ & $|\psi_{7b}\rangle^{\eqref{Eq.psi.num}}$ \\
6 & 7$\phantom{^a}$ & $|0\rangle^{\otimes 7}$ & 49$\phantom{^a}$ & $|\psi_7\rangle^{\eqref{psi7}}$ \\
7 & 1\textsuperscript{\hyperlink{FOOT}{c}} & \bcolor $|0\rangle^{\otimes 7}$ & 64\textsuperscript{\hyperlink{FOOT}{c}} & \bcolor $|\mathrm{GHZ}(7)\rangle$\\
\hline \hline
\multicolumn{5}{|c|}{$N = 8$} \\
\hline
1 & 0\textsuperscript{\hyperlink{FOOT}{a}} & \bcolor $|\mathrm{GHZ}(8)$ & 8\textsuperscript{\hyperlink{FOOT}{b}} & \bcolor $|0\rangle^{\otimes 8}$ 
\\
2 & 0$\phantom{^a}$ & \bcolor $|\psi_M\rangle_{12783456}^{\eqref{psi12783456}}$ & 28\textsuperscript{\hyperlink{FOOT}{b}} & \bcolor $|0\rangle^{\otimes 8}$ 
\\
3 & 0\textsuperscript{\hyperlink{FOOT}{a}} & \bcolor $|\mathrm{GHZ}(8)\rangle$ & 56\textsuperscript{\hyperlink{FOOT}{b}} & \bcolor $|0\rangle^{\otimes 8}$ 
\\
4 & 14$\phantom{^a}$ & $|\mathrm{tetra}\rangle^{\otimes 2}$ & 70$\phantom{^a}$ & $|0\rangle^{\otimes 8}$ 
\\
5 & 0\textsuperscript{\hyperlink{FOOT}{a}} & \bcolor $|\mathrm{GHZ}(8)\rangle$ & 90$\phantom{^a}$ & $|\mathrm{AME(6,2)}\rangle |0\rangle^{\otimes 2}$ 
\\
6 & 7$\phantom{^a}$ & $|\mathrm{GHZ}(7)\rangle |0\rangle$ & 168$\phantom{^a}$ & $|\psi^1_{12345678}\rangle^{\eqref{psi8_3uniform}}$ 
\\
7 & 0\textsuperscript{\hyperlink{FOOT}{a}} & \bcolor $|\mathrm{GHZ}(8)\rangle$ & 592/7$\phantom{^a}$ & $|\mathrm{tetra}(8)\rangle^{\eqref{tetra8}}$ 
\\
8 & 1\textsuperscript{\hyperlink{FOOT}{c}} & \bcolor $|0\rangle^{\otimes 8}$ & 129\textsuperscript{\hyperlink{FOOT}{b}} & \bcolor $|\mathrm{GHZ}(8)\rangle$ 
\\
\hline
\end{tabular}
\end{minipage}%
}
\caption{Examples of optimal $N$-qubit pure states that extremize the sector lengths $S_m$ [Eq.~\eqref{Eq.def.S}], presented in two tables: one sorted by $N$ (left) and the other by $m$ (right). Footnotes refer to either trivial (in)equalities or to previous work that identified and proved the optimality of some of these states. All states whose optimality is proven appear in a grey cell. They correspond to the extreme values of the region $\RR$ [see Eq.~\eqref{defR}], and therefore of the pure-states numerical range $\Scal$ [see Eq.~\eqref{defScal}]. The remaining states are conjectured optima, identified through algebraic-numerical methods. Some coincide with well-known states in the literature, but their optimality has not yet been rigorously proven.}

\label{tab.3}
\end{table*}
\begin{table*}[t]
\centering
\begin{tabular}{|c|c|@{\hspace{2pt}}c@{\hspace{2pt}}c|@{\;}c@{\;}|}
\hline
$\;N\;$ & $\;(N_A,N_{\bar{A}})\;$ & $\;\min [\Tr(\rho_A^2)+ R_{\rho_A}]$ & Optimal state & $(\Tr(\rho_A^2), R_{\rho_A})$ \\
\hline \hline
2 & (1,1) & $1$ & \bcolor{any $|\psi\rangle$} & $\;(x,1-x)$\footnote{with $x\in[1/2,1]$} \\
3 & (2,1) & $2^{-1}$ & \bcolor{$|0\rangle |\mathrm{GHZ}(2)\rangle$} & $(2^{-1},0)$ \\
4 & (3,1) & $2^{-1}$ & \bcolor{$|0\rangle |\mathrm{GHZ}(3)\rangle$} & $(2^{-1},0)$ \\
5 & (4,1) & $2^{-1}$ & \bcolor{$|0\rangle|\mathrm{GHZ}(4)\rangle$} & $(2^{-1},0)$ \\
6 & (5,1) & $2^{-1}$ & \bcolor{$|0\rangle|\mathrm{GHZ}(5)\rangle$} & $(2^{-1},0)$ \\
\hline \hline
3 & (1,2) & $1$ & any $|\psi\rangle$ & $(x,1-x)$\\
4 & (2,2) & $2^{-1}$ & $|0\rangle |\mathrm{GHZ}(3)\rangle$ & $(2^{-1},0)$ \\
5 & (3,2) & $2^{-2}$ & \bcolor{$|\mathrm{AME}(5,2)\rangle$} & $(2^{-2},0)$ \\
6 & (4,2) & $2^{-2}$ & \bcolor{$|0\rangle |\mathrm{AME}(5,2)\rangle$} & $(2^{-2},0)$ \\
7 & (5,2) & $2^{-2}$ & \bcolor{$\;|\mathrm{GHZ}(2)\rangle |\mathrm{AME}(5,2)\rangle\,$} & $(2^{-2},0)$ \\
8 & (6,2) & $2^{-2}$ & \bcolor{$|0\rangle^{\otimes 2} |\mathrm{AME}(6,2)\rangle$} & $(2^{-2},0)$ \\
9 & (7,2) & $2^{-2}$ & \bcolor{$|0\rangle^{\otimes 3} |\mathrm{AME}(6,2)\rangle$} & $(2^{-2},0)$ \\
\hline \hline
4 & (1,3) & $1$ & any $|\psi\rangle$ & $(x,1-x)$\\
5 & (2,3) & $2^{-1}$ & $|\mathrm{AME}(5,2)\rangle^\eqref{Eq.AME52.def}$  & $(2^{-2},2^{-2})$ \\
6 & (3,3) & $2^{-2}$ & $|\mathrm{AME}(6,2)\rangle^\eqref{Eq.AME62.def}$ & $(2^{-3},2^{-3})$ \\
7 & (4,3) & $2^{-3}$ & \bcolor{$|0\rangle|\mathrm{AME}(6,2)\rangle$} & $(2^{-3},0)$ \\
8 & (5,3) & $2^{-3}$ & \bcolor{$|0\rangle^{\otimes 2} |\mathrm{AME}(6,2)\rangle$} & $(2^{-3},0)$ \\
9 & (6,3) & $2^{-3}$ & \bcolor{$|0\rangle^{\otimes 3} |\mathrm{AME}(6,2)\rangle$} & $(2^{-3},0)$ \\
\hline \hline
5 & (1,4) & $1$ & any $|\psi\rangle$ & $(x,1-x)$\\
6 & (2,4) & $2^{-1}$ & $|\mathrm{AME}(6,2)\rangle$ & $(2^{-2},2^{-2})$ \\
7 & (3,4) & $2^{-2}$ & $|0\rangle |\mathrm{AME}(6,2)\rangle$ & $(2^{-2},0)$ \\
8 & (4,4) & $2^{-3}$ & $|0\rangle\left|\psi_M\right\rangle_{1234567}^{\eqref{psi1234567}}$
& $(2^{-3},0)$ \\
9 & (5,4) & $\mathit{2^{-4}}$ & \textit{numerical} & $\mathit{(2^{-4},0)}$ \\
10 & (6,4) & $\mathit{2^{-4}}$ & \textit{numerical} & $\mathit{(2^{-4},0)}$ \\
\hline \hline
6 & (1,5) & $1$ & any $|\psi\rangle$ & $(x,1-x)$\\
7 & (2,5) & $2^{-1}$ & $|\mathrm{AME}(6,2)\rangle |0\rangle$ & $(2^{-2},2^{-2})$ \\
8 & (3,5) & $2^{-2}$ & $|\mathrm{AME}(6,2)\rangle |0\rangle^{\otimes 2}$ & $(2^{-3},2^{-3})$ \\
9 & (4,5) & $2^{-3}$ & $|0\rangle|\psi_M\rangle_{12783456}^{\eqref{psi12783456}}$
& $(2^{-3},0)$ \\
10 & (5,5) & $\mathit{2^{-4}}$ & \textit{numerical} & $\mathit{(0.04787, 0.01463)}$ \\
\hline
\end{tabular}
\caption{Examples of states that minimize $\Tr(\rho_A^2)+ R_{\rho_A}$. For $N_A > N_{\bar{A}}$ (i.e., $k > N-k$), the minimum coincides with the lower bound $2^{-\min(k,N-k)}$ presented in Proposition~\ref{Prop.1}, which is therefore saturated by the corresponding optimal state. All states whose optimality is rigorously proven appear in a grey cell, the others being putative optimal states obtained by numerical optimization.
} 
\label{tab.4}
\end{table*}
\section{Proofs for the dual problem}
\subsection{Proof of feasibility Eq.~\texorpdfstring{\eqref{ineq_dual}}{Lg}}
\label{Qpositif1}
Here we show that the values given in \eqref{values_dual} provide a feasible point of the dual problem, namely that the inequalities $Q_m\geq c_m$ hold.

Let us first consider the special case $m=N$. Equation \eqref{defca}
gives $a_{qN}=0$ for all $q$, $a'_{kN}=0$ for $k\geq 1$ and $a'_{0N}=1/2^N$. We get
\begin{equation}    
\sum_{q=1}^{N/2-1} a_{qN}y_q+\sum_{k=0}^{N/2-1} a'_{k N}y'_k=1=c_N,
\end{equation}
thus Eq.~\eqref{ineq_dual} holds. For $m=1,...,N-1$, we only need to show that $Q_m\geq 0$. We have
    \begin{eqnarray}
    Q_m&=&\delta_{m\textrm{ even}}\sum_{q=1}^{N/2-1} \binom{N-m}{2q+1-m} y_q
    +\sum_{k=0}^{N/2-1}\left[\frac{1}{2^{N-k}}\binom{N-m}{k}-\frac{1}{2^{k}}\binom{N-m}{N-k}\right]y'_k\nonumber\\
    &=&\delta_{m\textrm{ even}}\left[2^{1-\frac{N}{2}}\binom{N-m}{\frac{N}{2}-m}\delta_{N\!\!\!\!\mod 4=2}+\!\!\sum_{q=\lfloor\frac{N+2}{4}\rfloor}^{\frac{N}{2}-1}\!\! \binom{N-m}{2q+1-m}2^{-2q+1}\right]
   +1\label{firstsum}\\
   &+&\!\!\!\!\!\!\sum_{k=1}^{\min(N-m,\frac{N}{2}-1)}\!\!\!\!\!\!\frac{1}{2^{N-k}}\binom{N-m}{k} \left[(-1)^k \left(2^{N-k}-2^k+1\right)-1\right]
   -\sum_{k=m}^{N/2-1}\frac{1}{2^{k}}\binom{N-m}{N-k} \left[(-1)^k \left(2^{N-k}-2^k+1\right)-1\right].\nonumber
\end{eqnarray}
By replacing $m$ by $r=N-m$ and setting $p =N-2 q -1$ in the first sum in \eqref{firstsum}, we get for $r=1,...,N-1$
\begin{eqnarray}
\label{qnr3}
   Q_{N-r} &=&\delta_{r\textrm{ even}}\left[2^{1-\frac{N}{2}}\binom{r}{\frac{N}{2}}\delta_{N\!\!\!\!\mod 4=2}+\!\!\sum_{\genfrac{}{}{0pt}{}{p=1}{p \textrm{ odd}}}^{2\lfloor\frac{N}{4}\rfloor-1}\!\!2^{p-N+2} \binom{r}{p}\right]+1\\
&+&\sum_{k=1}^{\min(r,\frac{N}{2}-1)}\frac{1}{2^{N-k}}\binom{r}{k} \left[(-1)^k \left(2^{N-k}-2^k+1\right)-1\right]-\sum_{k=N-r}^{N/2-1}\frac{1}{2^{k}}\binom{r}{N-k} \left[(-1)^k \left(2^{N-k}-2^k+1\right)-1\right].\nonumber
\label{Qnr}
\end{eqnarray}
The sums can be calculated analytically. If $r\leq N/2-1$, the last sum is 0 and Eq.~\eqref{qnr3} reduces to
\begin{equation}
\label{rsmall}
   Q_{N-r} =\delta_{r\textrm{ even}}\left[\sum_{\genfrac{}{}{0pt}{}{p=1}{p \textrm{ odd}}}^{r-1}\!\!2^{p-N+2} \binom{r}{p}\right]
   +1 +\sum_{k=1}^{r}\frac{1}{2^{N-k}}\binom{r}{k} \left[(-1)^k \left(2^{N-k}-2^k+1\right)-1\right].
\end{equation}
The last sum gives
\begin{equation}
\sum_{k=1}^{r}\frac{1}{2^{N-k}}\binom{r}{k} \left[(-1)^k \left(2^{N-k}-2^k+1\right)-1\right]=-1+\frac{1-(-3)^r+(-1)^r-3^r}{2^{N}},
\end{equation}
while
\begin{equation}
\label{identitesimple}
\sum_{\genfrac{}{}{0pt}{}{p=1}{p \textrm{ odd}}}^{r}\!\!2^{p} \binom{r}{p}=\frac{1}{2} \left(3^r-(-1)^r\right),
\end{equation}
hence \eqref{rsmall} becomes
\begin{equation}
\label{rsmall2}
   Q_{N-r} =\delta_{r\textrm{ even}}\left[\frac{3^r-(-1)^r}{2^{N-1}}\right]
  +\frac{1-(-3)^r+(-1)^r-3^r}{2^{N}} =0
\end{equation}
when considering in turn the cases $r$ even and odd.

 If $r\geq N/2$, the bounds of the sums are slightly different, and Eq.~\eqref{Qnr} becomes
\begin{eqnarray}
 Q_{N-r} &=&\delta_{r\textrm{ even}}\left[2^{1-\frac{N}{2}}\binom{r}{\frac{N}{2}}\delta_{N\!\!\!\!\mod 4=2}+\!\!\sum_{\genfrac{}{}{0pt}{}{p=1}{p \textrm{ odd}}}^{2\lfloor\frac{N}{4}\rfloor-1}\!\!2^{p-N+2} \binom{r}{p}\right]+1 \label{qnr2}\\
   &+&\sum_{k=1}^{\frac{N}{2}-1}\frac{1}{2^{N-k}}\binom{r}{k}\left[(-1)^k \left(2^{N-k}-2^k+1\right)-1\right]
   -\sum_{k=\frac{N}{2}+1}^{r}\frac{1}{2^{N-k}}\binom{r}{k} \left[(-1)^k \left(2^{k}-2^{N-k}+1\right)-1\right]
   \nonumber
\end{eqnarray}
where in the last sum we changed $k$ to $N-k$. In the last line , the terms in $(2^{N-k}-2^k)$ give
\begin{equation}
  \sum_{k=1}^{r}\frac{1}{2^{N-k}}\binom{r}{k} (-1)^k \left(2^{N-k}-2^k\right)=-1+\frac{1-(-3)^r}{2^{N}},
\end{equation}
while
\begin{eqnarray}
    \sum_{k=1}^{\frac{N}{2}-1}\frac{1}{2^{N-k}}\binom{r}{k} \left[(-1)^k -1\right]&=&-\frac{1}{2^{N-1}}\!\!\sum_{\genfrac{}{}{0pt}{}{k=1}{k \textrm{ odd}}}^{\frac{N}{2}-1}\!\!2^{k} \binom{r}{k}\\
    \sum_{k=\frac{N}{2}+1}^{r}\frac{1}{2^{N-k}}\binom{r}{k} \left[(-1)^k -1\right]&=&
    -\frac{1}{2^{N-1}}\!\!\sum_{\genfrac{}{}{0pt}{}{k=\frac{N}{2}+1}{k \textrm{ odd}}}^{r}\!\!2^{k} \binom{r}{k}.
\end{eqnarray}
The difference of the above two identities yields
\begin{eqnarray}
    \sum_{k=1}^{\frac{N}{2}-1}\frac{1}{2^{N-k}}\binom{r}{k} \left[(-1)^k -1\right]&-&\!\!\!
    \sum_{k=\frac{N}{2}+1}^{r}\frac{1}{2^{N-k}}\binom{r}{k} \left[(-1)^k -1\right]\\
    &=&
     -\frac{1}{2^{N-1}}\sum_{\genfrac{}{}{0pt}{}{k=1}{k \textrm{ odd}}}^{\frac{N}{2}-1}\!\!2^{k} \binom{r}{k}
 +\frac{1}{2^{N-1}}\sum_{\genfrac{}{}{0pt}{}{k=\frac{N}{2}+1}{k \textrm{ odd}}}^{r}\!\!2^{k} \binom{r}{k}\\
 &=&
     \frac{1}{2^{N-1}}\left(-2\sum_{\genfrac{}{}{0pt}{}{k=1}{k \textrm{ odd}}}^{\frac{N}{2}-1}\!\!2^{k} \binom{r}{k}
-2^{\frac{N}{2}} \binom{r}{\frac{N}{2}} \delta_{N\!\!\!\!\mod 4=2}
 +\sum_{\genfrac{}{}{0pt}{}{k=1}{k \textrm{ odd}}}^{r}\!\!2^{k} \binom{r}{k}\right)\,.\nonumber\\
\end{eqnarray}
We thus get
\begin{multline}
 Q_{N-r} =\delta_{r\textrm{ even}}\left[2^{1-\frac{N}{2}}\binom{r}{\frac{N}{2}}\delta_{N\!\!\!\!\mod 4=2}+\!\!\sum_{\genfrac{}{}{0pt}{}{k=1}{k \textrm{ odd}}}^{2\lfloor\frac{N}{4}\rfloor-1}\!\!2^{k-N+2} \binom{r}{k}\right]
 +\frac{1-(-3)^r}{2^{N}}\\ 
 -2^{2-N}\sum_{\genfrac{}{}{0pt}{}{k=1}{k \textrm{ odd}}}^{\frac{N}{2}-1}\!\!2^{k} \binom{r}{k}
-2^{1-\frac{N}{2}} \binom{r}{\frac{N}{2}} \delta_{N\!\!\!\!\mod 4=2}
 +\frac{3^r-(-1)^r}{2^{N}}
 \label{lastqnr}
\end{multline}
(using \eqref{identitesimple} for the last term).
For even $r$ and $N=4M+2\epsilon$, $\epsilon=0,1$, the upper bounds of the remaining sums are respectively $2M-1$ and $2M-1+\epsilon$, but since the sums run over odd $k$ only, the two sums cancel whether $\epsilon=0$ or 1; thus $Q_{N-r}=0$ for even $r$. For odd $r$, Eq.~\eqref{lastqnr} reduces to
\begin{eqnarray}
    Q_{N-r} &=&\frac{1}{2^{N-2}}\left(\frac{1+3^r}{2}
 -\sum_{\genfrac{}{}{0pt}{}{k=1}{k \textrm{ odd}}}^{\frac{N}{2}-1}\!\!2^{k} \binom{r}{k}-2^{\frac{N}{2}-1} \binom{r}{\frac{N}{2}} \delta_{N\!\!\!\!\mod 4=2}\right)\nonumber\\
&=&\frac{1}{2^{N-2}}\left(\sum_{\genfrac{}{}{0pt}{}{k=\frac{N}{2}}{k \textrm{ odd}}}^{r}\!\!2^{k} \binom{r}{k}
-\frac12 2^{\frac{N}{2}} \binom{r}{\frac{N}{2}} \delta_{N\!\!\!\!\mod 4=2}\right)\,,
\end{eqnarray}
where we used again Eq.~\eqref{identitesimple}. It is a sum of positive terms, apart from the last term, which, for $N/2$ odd, is half the term corresponding to $k=N/2$ in the sum. Therefore $Q_{N-r}\geq 0$. 

\subsection{Proof of Eq.~\texorpdfstring{\eqref{eq_min}}{Lg}}
\label{Qpositif2}

We have
\begin{equation}
    b_{q}=\left(2^{2q}-1\right)\mbinom{N}{2q+1},\qquad
    b'_k=\frac{2^{N-k}-2^k}{2^N}\mbinom{N}{k}.
\end{equation}
Using the values \eqref{values_dual} for $y_q,y'_k$, we get
\begin{eqnarray}
 \sum_{q=1}^{N/2-1} b_q y_q+\sum_{k=0}^{N/2-1} b'_k y'_k
  &=&\left(1-2^{1-\frac{N}{2}}\right)\mbinom{N}{\frac{N}{2}}\delta_{N\!\!\!\!\mod 4=2}+2\!\!\sum_{q=\lfloor\frac{N+2}{4}\rfloor}^{\frac{N}{2}-1}\!\! \left(1-2^{-2q}\right)\mbinom{N}{2q+1}\nonumber\\
   &+&2^{N}-1+\frac{1}{2^N}\sum_{k=1}^{\frac{N}{2}-1}\left(2^{N-k}-2^k\right)\mbinom{N}{k}\left[(-1)^k \left(2^{N-k}-2^k+1\right)-1\right].
\end{eqnarray}
The first sum is
\begin{equation}
\label{sum1}
    2\!\!\sum_{q=\lfloor\frac{N+2}{4}\rfloor}^{\frac{N}{2}-1}\!\! \left(1-2^{-2q}\right)\mbinom{N}{2q+1}= 2\!\!\sum_{\genfrac{}{}{0pt}{}{k=1}{k \textrm{ odd}}}^{\frac{N}{2}-1}\!\! \left(1-2^{-N+k+1}\right)\mbinom{N}{k},
\end{equation}
while the second sum reads
\begin{equation}
\label{sum2}
   \sum_{\genfrac{}{}{0pt}{}{k=1}{k \textrm{ odd}}}^{\frac{N}{2}-1}\mbinom{N}{k}\left(2^{-k}-2^{k-N}\right)\left(2^k-2^{N-k}-2\right)+\frac{1}{2^N}  \sum_{\genfrac{}{}{0pt}{}{k=2}{k \textrm{ odd}}}^{\frac{N}{2}-1}\mbinom{N}{k}\left(2^{N-k}-2^k\right)^2\,.
\end{equation}
Summing up Eqs.~\eqref{sum1} and \eqref{sum2} gives
\begin{equation}
\label{eqg}
\!\!\sum_{\genfrac{}{}{0pt}{}{k=1}{k \textrm{ odd}}}^{\frac{N}{2}-1}\!\! \mbinom{N}{k}\left(4-2^{-N+k+1}-2^{2k-N}-2^{N-2k}-2^{1-k}\right)+\frac{1}{2^N}  \sum_{\genfrac{}{}{0pt}{}{k=2}{k \textrm{ even}}}^{\frac{N}{2}-1}\mbinom{N}{k}\left(2^{N-k}-2^k\right)^2.
\end{equation}
For the second sum we get
\begin{eqnarray}
        \sum_{\genfrac{}{}{0pt}{}{k=2}{k \textrm{ even}}}^{\frac{N}{2}-1}\mbinom{N}{k}\left(2^{N-k}-2^k\right)^2&=&\sum_{\genfrac{}{}{0pt}{}{k=2}{k \textrm{ even}}}^{N}\mbinom{N}{k}\left(2^{N-k}-2^k\right)^2-  \sum_{\genfrac{}{}{0pt}{}{k=\frac{N}{2}}{k \textrm{ even}}}^{N}\mbinom{N}{k}\left(2^{N-k}-2^k\right)^2 \\
        &=&\sum_{\genfrac{}{}{0pt}{}{k=2}{k \textrm{ even}}}^{N}\mbinom{N}{k}\left(2^{N-k}-2^k\right)^2-\sum_{\genfrac{}{}{0pt}{}{k=0}{k \textrm{ even}}}^{\frac{N}{2}}\mbinom{N}{k}\left(2^{k}-2^{N-k}\right)^2 
\end{eqnarray}
and thus
\begin{equation}
     \sum_{\genfrac{}{}{0pt}{}{k=2}{k \textrm{ even}}}^{\frac{N}{2}-1}\mbinom{N}{k}\left(2^{N-k}-2^k\right)^2=\frac12 \sum_{\genfrac{}{}{0pt}{}{k=2}{k \textrm{ even}}}^{N}\mbinom{N}{k}\left(2^{N-k}-2^k\right)^2-\left(1-2^{N}\right)^2 =\frac{5^N+3^N-3\times 4^{N}+2^{N+2}-2}{2} .
\end{equation}
For the odd terms in \eqref{eqg}, and in the same way changing $k$ to $N-k$, we obtain
\begin{equation}
    \!\!\sum_{\genfrac{}{}{0pt}{}{k=1}{k \textrm{ odd}}}^{\frac{N}{2}-1}\!\! \mbinom{N}{k}\left(4-2^{-N+k+1}-2^{2k-N}-2^{N-2k}-2^{1-k}\right)=\!\!\sum_{\genfrac{}{}{0pt}{}{k=\frac{N}{2}+1}{k \textrm{ odd}}}^{N-1}\!\! \mbinom{N}{k}\left(4-2^{-k+1}-2^{N-2k}-2^{2k-N}-2^{1-N+k}\right) .
\end{equation}
Thus, adding both sides and dividing by 2,
\begin{multline}
        \sum_{\genfrac{}{}{0pt}{}{k=1}{k \textrm{ odd}}}^{\frac{N}{2}-1}\!\! \mbinom{N}{k}\left(4-2^{-N+k+1}-2^{2k-N}-2^{N-2k}-2^{1-k}\right)\\
=\frac12
    \sum_{\genfrac{}{}{0pt}{}{k=1}{k \textrm{ odd}}}^{N-1}\!\! \mbinom{N}{k}\left(4-2^{-k+1}-2^{N-2k}-2^{2k-N}-2^{1-N+k}\right) - \mbinom{N}{\frac{N}{2}}(1-2^{-\frac{N}{2}+1})\delta_{N\!\!\!\!\mod 4=2}\nonumber\\
    =\frac{2^{2N+1}-3^N-5^N+2}{2^{N+1}}- \mbinom{N}{\frac{N}{2}}(1-2^{-\frac{N}{2}+1})\delta_{N\!\!\!\!\mod 4=2}\,.\hspace{6.4cm}
\end{multline}
Gathering together all identities, Eq.~\eqref{eq_min} yields
\begin{eqnarray}
\label{eq_min2}
\sum_{q=1}^{N/2-1} b_q y_q+\sum_{k=0}^{N/2-1} b'_k y'_k
  &=&2^{N}-1+\frac{5^N+3^N-3\times 4^{N}+2^{N+2}-2}{2^{N+1}}+\frac{2^{2N+1}-3^N-5^N+2}{2^{N+1}}\\
 &=&2^{N-1}+1,
 \end{eqnarray}
which completes the proof.
\end{widetext}

\end{appendix}
\bibliographystyle{apsrev4-2}
\bibliography{Refs}
\end{document}